   \newcommand{\be}[0]{\begin{equation}}
   \newcommand{\ee}[0]{\end{equation}}
   \newcommand{\ba}[0]{\begin{eqnarray}}
   \newcommand{\ea}[0]{\end{eqnarray}}  

	\newcommand{\MS}[0]{\overline{{\rm MS}} }
	\newcommand{\RS}[0]{\overline{{\rm RS}} }
\documentstyle[12pt,epsfig]{article}

\begin{document}

\topmargin=-0.5in
\headheight=0.6in        
\headsep=0in
\footheight=0.5in
\footskip=0.5in

\begin{titlepage}

\Large
\hfill\vbox{\hbox{DTP/96/52}
            \hbox{June 1996}}
\nopagebreak

\vspace{0.75cm}
\begin{center}
\LARGE
{\bf RS-Invariant All-Orders Renormalon Resummations for some QCD Observables}
\vspace{0.6cm}
\Large

C.J.Maxwell and D.G.Tonge

\vspace{0.4cm}
\large
\begin{em}
Centre for Particle Theory, University of Durham\\
South Road, Durham, DH1 3LE, UK
\end{em}

\vspace{1.7cm}

\end{center}
\normalsize
\vspace{0.45cm}

\begin{abstract}

We propose a renormalon-inspired resummation of QCD perturbation theory 
based on approximating the renormalization scheme (RS) invariant effective 
charge beta-function coefficients by the portion containing the
highest power of 
$b$=$\frac{1}{6}(11N$--$2N_{f})$, for SU($N$) QCD with 
$N_{f}$ quark flavours. This can be accomplished using exact large-$N_{f}$ 
all-orders results. The resulting resummation is RS-invariant and the exact 
next-to-leading order (NLO) and next-to-NLO (NNLO) coefficients in any RS are 
included. This improves on a previously employed naive resummation of the 
leading-$b$ piece of the perturbative coefficients which is
RS-dependent, making 
its comparison with fixed-order perturbative results ambiguous. The
RS-invariant 
resummation is used to assess the reliability of fixed-order
perturbation theory 
for the $e^{+}e^{-}$ $R$-ratio, the analogous $\tau$-lepton decay 
ratio $R_{\tau}$, and Deep Inelastic Scattering (DIS) sum rules, by
comparing it 
with the exact NNLO results in the effective charge RS. For the $R$-ratio and 
$R_{\tau}$, where large-order perturbative behaviour is dominated by a leading 
ultra-violet renormalon singularity, the comparison indicates fixed-order 
perturbation theory to be very reliable. For DIS sum rules, which have
a leading 
infra-red renormalon singularity, the performance is rather poor. In
this way we 
estimate that at LEP/SLD energies ideal data on the $R$-ratio could determine 
$\alpha_{s}(M_{Z})$ to three-significant figures, and for the $R_{\tau}$ 
we estimate a theoretical uncertainty $\delta\alpha_{s}(m_{\tau})\simeq0.008$ 
corresponding to $\delta\alpha_{s}(M_{Z})\simeq0.001$. This encouragingly small 
uncertainty is much less than has recently been deduced from
comparison with the 
ambiguous naive resummation.
\end{abstract}

\end{titlepage}

\newpage

\section{Introduction}

There has been a great deal of interest recently in the possibility of 
identifying and resumming to all-orders the Feynman diagrams which dominate 
the 
large-order asymptotics of perturbation theory [1--6].

More precisely consider some generic dimensionless QCD observable $D(Q)$, 
dependent on the single dimensionful scale $Q$, with a perturbation series
\be
D(Q)=a+d_{1}a^{2}+d_{2}a^{3}+\cdots+d_{k}a^{k+1}+\cdots\;, 
\ee 
where
$a\equiv\alpha_{s}/\pi$ is the renormalization group (RG) improved
coupling. The perturbative coefficients $d_{k}$ can themselves be
written as polynomials of degree $k$ in the number of quark flavours,
$N_{f}$; we shall assume massless quarks. 
\be 
d_{k}=d_{k}^{[k]}N_{f}^{k}+d_{k}^{[k-1]}N_{f}^{k-1}+
\cdots+d_{k}^{[0]}\;.  
\ee 
The leading $d_{k}^{[k]}$ coefficient corresponds to the evaluation in each 
order 
of perturbation theory of a gauge-invariant set of Feynman diagrams containing 
chains of $k$ fermion bubbles. Techniques for evaluating such diagrams exactly 
in 
all-orders have been developed, and rather compact results for QCD vacuum 
polarization \cite {ben1,broad}, Deep Inelastic Scattering (DIS) sum rules 
\cite {broadkat}, and heavy 
quark 
decay widths and pole masses \cite {benbraun2,bert4} obtained.

The resummation of such diagrams provides a gauge-invariant effective charge in 
QED, but in QCD one would need to include chains of gluon bubbles and ghosts, 
and 
the isolation of a gauge-invariant  subset of diagrams providing an analogous 
QCD 
effective charge is problematic \cite {watson}. One knows on the grounds of 
gauge 
invariance, however, that part of the result should be proportional to $b^{k}$, 
where $b$ is the first beta-function coefficient, 
$b$=$\frac{1}{6}(11N$--$2N_{f})$, for SU($N$) QCD. Since for large-$N_{f}$ one 
must obtain the QED result one can substitute $N_{f}$=$(\frac{11}{2}N-3b)$ in 
the
`$N_{f}$-expansion' of equation (2) to obtain a `$b$-expansion' 
\be
d_{k}=d_{k}^{(k)}b^{k}+d_{k}^{(k-1)}b^{k-1}+\cdots+d_{k}^{(0)}\;,
\ee
where
$d_{k}^{[k]}$=$(-1/3)^{k}d_{k}^{(k)}$, so that exact knowledge of the 
leading-$N_{f}$ $d_{k}^{[k]}$ to all-orders implies exact knowledge
of the leading-$b$ $d_{k}^{(k)}$ to all-orders as well.

The existence of so-called infra-red (IR) and ultra-violet (UV) renormalon 
singularities in the Borel plane at positions $z$=$2l/b$ with $l$ a positive or 
negative integer, respectively, means that in large-orders the perturbative 
coefficients are expected to grow as $d_{k}\sim b^{k}k!$. As shown in reference 
\cite {charles1} this singularity structure leads to the expectation that the 
`leading-$b$' 
term 
when expanded in powers of $N_{f}$ should, asymptotically, reproduce the 
sub-leading coefficients. That is, expressing $d_{k}^{(k)}b^{k}$ as 
\be 
d_{k}^{(k)}b^{k}=d_{k}^{[k]}N_{f}^{k}+\tilde{d}_{k}^{[k-1]}N_{f}^{k-1}+
\cdots+\tilde{d}_{k}^{[k-r]}N_{f}^{k-r}+\cdots+\tilde{d}_{k}^{[0]}\;,  
\ee 
one can show that,
\be
\tilde{d}_{k}^{[k-r]}\sim d_{k}^{[k-r]}\left[1+O(\frac{1}{k})\right]\;,
\ee
so that for fixed $r$ and large $k$ the sub-leading `$N_{f}$-expansion' 
coefficients are reproduced.

As demonstrated in reference \cite {charles2} for both the $e^{+}e^{-}$ Adler 
$D$-function 
and DIS sum rules this asymptotic dominance of the 
leading-$b$ term is already apparent in comparisons with the exact 
next-to-leading 
order (NLO) and next-to-NLO (NNLO) perturbative coefficients, $d_{1}$ and 
$d_{2}$. 
For instance for SU($N$) QCD the first two perturbative coefficients for the 
Adler 
$D$-function are given by \cite {kat,gorish1}
\ba
d_{1}&=&-.115N_{f}+\left(.655N+\frac{.063}{N}\right) \;,\\
d_{2}&=&.086N_{f}^{2}+N_{f}\left(-1.40N-\frac{.024}{N}\right)
+\left(2.10N^{2}-.661-\frac{.180}{N^{2}}\right)
\;.
\ea
These are to be compared with the leading-$b$ terms
\ba
d_{1}^{(1)}b&=&.345b=-.115N_{f}+.634N\;,\\
d_{2}^{(2)}b^{2}&=&.776b^{2}=.086N_{f}^{2}-.948N_{f}N+2.61N^{2}\;.
\ea
The subleading $N$ and $N_{f}N$, $N^{2}$ coefficients approximate 
well in sign and magnitude those in the exact expressions in equations
(6) and (7).

Notice that the level of accuracy with which the sub-leading coefficients 
$d_{k}^{[k-r]}$ are reproduced is far in excess of that to be anticipated from 
the 
asymptotic expectation of equation (5). This is a rather weak statement which 
implies only that $d_{k}^{[k-r]}$ should be reproduced to $O(1/k)$ accuracy for 
fixed $r$ and large $k$ on expanding $d_{k}^{(k)}b^{k}$, whereas the 
$d_{k}^{[0]}$ 
($r$=$k$) coefficient, which is leading in the $1/N$ expansion for a large 
number 
of colours, is reproduced accurately for $k$=1 and $k$=2.

The above observations suggest that there should be some merit in resumming to 
all-orders the `leading-$b$' terms, even though many features of the 
approximation 
remain to be clarified. In a number of recent papers [1--6] such a
programme has 
been carried out. We shall refer mainly to the results of reference \cite 
{charles2}.

Following \cite {charles2} we define
\be
D^{(L)}\equiv\sum_{k=0}^{\infty}d_{k}^{(L)}a^{k+1}\;
\ee
where
$d_{k}^{(L)}\equiv d_{k}^{(k)}b^{k}$ ($d_{0}^{(L)}\equiv 1$). We can also 
consider the 
complementary sum over the sub-leading $b$ terms
\be
D^{(NL)}\equiv\sum_{k=1}^{\infty}d_{k}^{(NL)}a^{k+1}\;
\ee
where
$d_{k}^{(NL)}\equiv d_{k}-d_{k}^{(L)}$. Hence  
\[
D=D^{(L)}+D^{(NL)}\nonumber\;.
\]

In reference \cite {charles2} the summation in equation (10) was defined using 
Borel 
summation, with the IR renormalon singularities principal value (P.V.) 
regulated. 
Whilst such a summation can be performed the results for $D^{(L)}$ and 
$D^{(NL)}$ are separately dependent on the chosen renormalization scheme (RS), 
the sum of the two being formally RS-invariant. Changing the RS changes the 
definition of the renormalization group (RG)-improved coupling and hence the 
`$a$' appearing in the summation changes. In the full sum this change is 
precisely compensated by the RG transformations of the coefficient $d_{k}$ 
under 
the change in RS. However, by restricting oneself to the `leading-$b$' piece of 
the coefficients this exact compensation is destroyed and the resulting sum 
$D^{(L)}$ is fatally RS-dependent \cite {charles2,chyla}. One response to this, 
which has been adopted 
in references [2--6], is to artificially restrict the
RG-transformation of `$a$' 
to that contributed by the first term in the beta-function only. With this 
restriction $D^{(L)}$ is then RS-invariant since RS changes are exactly 
compensated for at the `leading-$b$' level. In our view, however, one can and 
should do much better than this. For several QCD observables one has exact 
results for the first two perturbative coefficients $d_{1}$ and $d_{2}$, 
usually 
in the modified minimal subtraction ($\MS$) scheme with renormalization scale 
$\mu$=$Q$ \cite {kat,gorish1,gorish2,larin}. What is clearly needed is a 
resummation 
in which the 
exact 
NLO 
and NNLO contributions are included, and an approximate resummation of the 
higher orders performed, in such a way that the full sum is explicitly 
RS-invariant under the full QCD RG transformations. In this way one would have 
a 
test bed for assessing the reliability of fixed-order perturbation
theory in any 
RS by seeing how it differed from the RS-invariant resummed result. As one 
reduced the energy scale $Q$ (e.g. the centre of mass-energy in $e^{+}e^{-}$ 
annihilation) one could also assess how the reliability of fixed-order 
perturbation theory deteriorates.

In this paper we shall show in section 2 how such an improved resummation can 
be 
carried out, and will use it in section 3 to assess the reliability of 
fixed-order perturbation theory for the $e^{+}e^{-}$ $R$-ratio, the analogous 
$\tau$-decay ratio $R_{\tau}$, and DIS sum rules. Section 4 contains discussion 
of results and section 5 our conclusions.

\section{RS Dependence and RS-Invariants}

We begin by briefly reviewing the RS dependence of the coupling `$a$' and the 
perturbative coefficients $d_{k}$. We refer the reader to reference \cite 
{reader} for 
more details.

The RG improved coupling `$a$' satisfies the beta-function equation
\be
\frac{\mbox{d}a}{\mbox{d}\ln\tau}=-a^{2}(1+ca+c_{2}a^{2}+\cdots+
c_{k}a^{k}+\cdots)\equiv -\beta(a)
\;,
\ee
where
$\tau\equiv b\ln\frac{\mu}{\Lambda}$, with $\mu$ the renormalization scale and 
$\Lambda$ the dimensional transmutation mass parameter of QCD. Here $b$ and $c$ 
are universal with
\ba
b&=&\frac{1}{6}(11C_{A}-2N_{f})\;,\nonumber\\
c&=&\left[-\frac{7}{8}\frac{C_{A}^{2}}{b}-
\frac{11}{8}\frac{C_{A}C_{F}}{b}+\frac{5}{4}C_{A}
+\frac{3}{4}C_{F}\right]\;,
\ea
where for SU($N$) QCD $C_{A}$=$N$ and $C_{F}$=$(N^{2}$--$1)/2N$.

As shown by Stevenson \cite {stev1} the RS is labelled by the
parameters $\tau$, 
$c_{2}$, $c_{3}$, $\ldots$; the higher beta-function coefficients are not 
universal and characterise the RS. Integrating up equation (12) with a suitable 
choice of boundary condition \cite {stev1} one finds that 
$a(\tau,c_{2},c_{3},\cdots,c_{k})$ is the solution of the transcendental 
equation 
\be
\tau=\frac{1}{a}+c\ln\frac{ca}{1+ca}+
\int_{0}^{a}\mbox{d}x\left[-\frac{1}{x^{2}B(x)}+\frac{1}{x^{2}(1+cx)}
\right]\;,
\ee
where $B(x)\equiv (1+cx+c_{2}x^{2}+c_{3}x^{3}+\cdots+c_{k}x^{k}+\cdots)$.

For consistency of perturbation theory one finds that $d_{1}(\tau)$, 
$d_{2}(\tau,c_{2})$, $d_{3}(\tau,c_{2},c_{3})$, $\cdots$, 
$d_{k}(\tau,c_{2},c_{3},\cdots,c_{k})$, $\cdots$. The combination
\be
\rho_{0}(Q)=\tau-d_{1}(\tau)\equiv b\ln\frac{Q}{\overline \Lambda}\;
\ee
is RS-invariant.

The explicit functional dependence of the $d_{k}$ on RS is conveniently 
obtained by considering the special RS, the effective charge (EC) scheme \cite 
{grun1}, 
in which $d_{1}$=$d_{2}$=$\cdots$=$d_{k}$=$\cdots$=$0$, so that $D$=$a$, and in 
this scheme the renormalized coupling is the observable itself. From equation 
(15) this RS will correspond to the choice of parameter $\tau$=$\rho_{0}$ 
(ensuring $d_{1}$=$0$). To determine the remaining parameters, 
$c_{2}$=$\rho_{2}$, $c_{3}$=$\rho_{3}$, $\cdots$, $c_{k}$=$\rho_{k}$, $\cdots$, 
characterizing the EC RS, one proceeds as follows. From equation (12) we have 
that for two RS's, barred and unbarred,
\be
\overline \beta(\overline a)=\frac{\mbox{d}\overline 
a}{\mbox{d}a}\beta(a(\overline a))\;,
\ee
where
$a$, $\overline a$ denote the couplings in the respective schemes RS and $\RS$. 
If the barred RS is chosen to be the EC scheme then
\be
\overline \beta(\overline a)=\rho(\overline a)=\overline a^{2}(1+c\overline a+ 
\rho_2\overline a^{2}+\cdots+\rho_{k}\overline a^{k}+\cdots)\;,
\ee
with $\overline a$=$D$. Then equation (16) gives
\be
\rho(D)=\frac{\mbox{d}D}{\mbox{d}a}\beta(a(D))\;,
\ee
where
$a(D)$ is the inverted perturbation series.

By expanding both sides of equation (18) as power series in $D$ and equating 
coefficients one obtains
\begin{eqnarray}
\rho_{2}&=&c_{2}+d_{2}-cd_{1}-d_{1}^2 \nonumber\\
\rho_{3}&=&c_{3}+2d_{3}-4d_{1}d_{2}-2d_{1}\rho_{2}-cd_{1}^{2}+2d_{1}^{3} \\
\vdots & & \vdots \nonumber 
\end{eqnarray}
Since $\rho_{0}$=$\tau-d_{1}$ is RS-invariant we can use $d_{1}$ itself, rather 
than $\tau$, to label the RS. Rearranging equation (19) we can then obtain
\begin{eqnarray}
d_{2}(d_{1},c_{2})&=&d_{1}^2+cd_{1}+(\rho_{2}-c_{2}) \nonumber\\
d_{3}(d_{1},c_{2},c_{3})&=&d_{1}^{3} + 
\frac{5}{2}cd_{1}^{2}+(3\rho_{2}-2c_{2})d_{1}+\frac{1}{2}(\rho_{3}-c_{3}) \\
\vdots & & \vdots \nonumber 
\end{eqnarray}
The result for $d_{n}(d_{1},c_{2},\cdots,c_{n})$ is a polynomial of degree 
$n$ in $d_{1}$ with coefficients involving $\rho_{n}$,$\rho_{n-1}$,$\cdots$,$c$ 
and $c_{2}$,$c_{3}$,$\cdots$,$c_{n}$; such that 
$d_{n}$(0,$\rho_{2}$,$\rho_{3}$,$\cdots$,$\rho_{n}$)=0. The 
$\rho_{2}$,$\rho_{3}$,$\cdots$,$\rho_{n}$,$\cdots$, are RS-invariants which 
completely characterise the QCD observable $D$. They are independent of the 
energy scale $Q$ but do depend on the number of active quark flavours, $N_{f}$. 
Given just these numbers the perturbative coefficients in any RS can
be obtained 
from equations (20). To construct RS-invariant resummations the
strategy will be 
to approximate the RS-invariants $\rho_{k}$, and then use equations (20) to 
obtain the approximate perturbative coefficients in any arbitrary RS. In this 
way invariance under the full RG transformations of QCD is guaranteed.

The $\rho_{k}$ invariants can be organised as an expansion in $b$, with
\be
\rho_{k}=\rho_{k}^{(k)}b^{k}+\rho_{k}^{(k-1)}b^{k-1}+\cdots+\rho_{k}^{(0)}+ 
\rho_{k}^{(-1)}b^{-1}\;.
\ee
The $b^{-1}$ term arises from the fact that in a `regular' RS such as minimal 
subtraction the $d_{k}$ are polynomials in $b$ of degree $k$ \cite {grun2}, 
whereas the 
corresponding beta-function coefficients $c_{k}$ are polynomials in $b$ of 
degree $k$-1 with additional $b^{-1}$ terms (c.f. the expression for
$c$=$c_{1}$ 
in equation (13)). The RS-invariant combinations in equation (19) in principle 
could contain arbitrary inverse powers of $b$, but RG considerations guarantee 
that only $b^{-1}$ terms remain \cite {grun2}. Thus $b\rho_{k}$ is a polynomial 
of degree 
$k$+1 in $b$.

The effective charge beta-function $\rho(D)$ (equation (17))
will contain Borel plane singularities at the same positions as those in $D(a)$ 
\cite {ben2} and hence one should expect a weak asymptotic result analogous to 
equation 
(5), with the $\rho_{k}^{(k)}b^{k+1}$ term asymptotically reproducing the 
$N_{f}$-expansion coefficients of $b\rho_{k}$. For the Adler D-function and DIS 
sum rules the level at which this works is again far in excess of that to be 
anticipated from the asymptotic result. The $\rho_{k}^{(k)}$ term involves only 
combinations of the $d_{k}^{(k)}$, with for instance 
$\rho_{2}^{(2)}$=$d_{2}^{(2)}-(d_{1}^{(1)})^{2}$, and so the $\rho_{k}^{(k)}$ 
can be 
obtained to all-orders given the exact leading-$N_{f}$ all-orders results.

For the Adler D-function ($\tilde{D}$ \cite {charles2}) one has the
exact result 
for SU($N$) 
QCD (where the ``light-by-light'' contribution $\tilde{\tilde{D}}$ is excluded, 
see \cite {charles2})
\ba
b\rho_{2}(\tilde{D})&=&-0.0243N_{f}^{3}+(0.553N-0.00151\frac{1}{N})N_{f}^{2}
\nonumber\\
& 
&+(-3.32N^{2}+0.344+0.0612\frac{1}{N^{2}})N_{f}\\
& &+(3.79N^{3}-1.45N-0.337\frac{1}{N})\nonumber
\;.
\ea
This is to be compared with the `leading-$b$' piece
\ba
b^{3}\rho_{2}^{(2)}(\tilde{D})&=&b^{3}(d_{2}^{(2)}-(d_{1}^{(1)})^{2})=0.656b^{3}
\nonumber\\
&=&-0.0243N_{f}^{3}+0.401NN_{f}^{2}-2.21N^{2}N_{f}+4.04N^{3}
\;.
\ea
Notice the good agreement of the sub-leading $NN_{f}^{2}$, $N^{2}N_{f}$, and 
$N^{3}$ coefficients.

For the DIS sum rules (polarized Bjorken or GLS, $\tilde{K}$ \cite {charles2}) 
one has the exact 
result
\ba
b\rho_{2}(\tilde{K})&=&-0.0221N_{f}^{3}+(0.513N+0.00665\frac{1}{N})N_{f}^{2}
\nonumber\\
& &+(-3.29N^{2}+0.505+0.0143\frac{1}{N^{2}})N_{f}\\
& &+(3.85N^{3}-1.73N-0.337\frac{1}{N})\nonumber
\;,
\ea
which is to be compared with
\ba
b^{3}\rho_{2}^{(2)}(\tilde{K})&=&0.597b^{3}
\nonumber\\
&=&-0.0221N_{f}^{3}+0.365NN_{f}^{2}-2.01N^{2}N_{f}+3.68N^{3}
\;,
\ea
again the $NN_{f}^{2}$, $N^{2}N_{f}$, and $N^{3}$ coefficients are well 
reproduced.

In both cases in the large-$N$ limit ($N_{f}$=0) the RS-invariant $\rho_{2}$ is 
approximated to better than 10\% accuracy. The 20\% level agreement 
of the sub-leading coefficients does not, unfortunately, guarantee that the 
overall RS-invariant is reproduced to the same accuracy for all $N$, $N_{f}$ 
since there are large numerical cancellations. For instance for $N$=3 and 
$N_{f}$=5 one has $\rho_{2}(\tilde{D})$=-2.98 (exact), whereas 
$b^{2}\rho_{2}^{(2)}(\tilde{D})$=9.64.

\section{RS-Invariant Leading-$b$ Resummations}

\setcounter{figure}{0}
\renewcommand{\thefigure}{1(\alph{figure})}

Approximating the RS-invariants $\rho_{k}$ by 
$\rho_{k}^{(L)}\equiv b^{k}\rho_{k}^{(k)}$ one can now define the RS-invariant 
resummation
\be
D^{(L*)}\equiv\sum_{k=0}^{\infty}d_{k}^{(L*)}a^{k+1}\;,
\ee
where in a general RS $d_{k}^{(L*)}(d_{1},c_{2},c_{3},\cdots,c_{k})$
is obtained 
by replacing $\rho_{k}$ in equations (20) by $\rho_{k}^{(L)}$, so that 
\ba
d_{0}^{(L*)}&=&1\nonumber\\
d_{1}^{(L*)}&=&d_{1}\nonumber\\
d_{2}^{(L*)}(d_{1},c_{2})&=&d_{1}^2+cd_{1}+(\rho_{2}^{(L)}-c_{2}) \\
d_{3}^{(L*)}(d_{1},c_{2},c_{3})&=&d_{1}^{3} + 
\frac{5}{2}cd_{1}^{2}+(3\rho_{2}^{(L)}-2c_{2})d_{1}+
\frac{1}{2}(\rho_{3}^{(L)}-c_{3}) 
\nonumber\\
\vdots & & \vdots \nonumber 
\;
\ea
Notice that, unlike the strict `leading-$b$' approximation of equation
(10), the 
NLO coefficient $d_{1}$ is now included exactly. If an exact NNLO calculation 
exists then the exact $\rho_{2}$ can be used and $\rho_{3}$,$\rho_{4}$,$\cdots$ 
approximated by $\rho_{3}^{(L)}$,$\rho_{4}^{(L)}$,$\cdots$, so that $d_{2}$ (in 
any RS) is included exactly. In any case the all-orders sum in equation (26) is 
formally RS-invariant, and can be compared with the NLO, NNLO, ${\rm 
N}^{3}$LO,$\cdots$,${\rm N}^{n}$LO,$\cdots$ fixed-order perturbative 
approximations to assess the accuracy of the fixed-order results,
\be
D^{(L*)(n)}\equiv\sum_{k=0}^{n}d_{k}^{(L*)}a^{k+1}\;.
\ee

The next task is to define the all-orders resummation in equation (26). If we 
consider the EC RS then $a$=$D$ and
\[
\tau=\rho_{0}=b\ln\frac{Q}{\Lambda_{\MS}}-d_{1}^{\MS}(\mu=Q)\nonumber\;,
\]
where for later convenience we have expressed the RS invariant $\rho_{0}$ in 
terms of the $\MS$ scheme NLO coefficient with $\mu$=$Q$, $d_{1}^{\MS}(\mu=Q)$, 
which is customarily what is computed, and where $\Lambda_{\MS}$ is the 
universal dimensional transmutation parameter of QCD. Equation (14) in the EC 
scheme with $x^{2}B(x)$=$\rho(x)$, the EC beta-function of equation (17), then 
yields
\be
\frac{1}{D}+c\ln\frac{cD}{1+cD}=b\ln\frac{Q}{\Lambda_{\MS}}-d_{1}^{\MS}(\mu=Q)-
\int_{0}^{D}\mbox{d}x\left[-\frac{1}{\rho(x)}+
\frac{1}{x^{2}(1+cx)}\right]\;.
\ee
As discussed in reference \cite {reader} the EC beta-function $\rho(x)$ is of 
fundamental 
significance since
\be
\frac{\mbox{d}D(Q)}{\mbox{d}\ln{Q}}=-b\rho(D(Q))\;,
\ee
and so it can be partially reconstructed from measurements of the energy 
evolution of the observable. Given $\rho(x)$, $D(Q)$ is specified by the 
solution of the transcendental equation (29). The resummation $D^{(L*)}$ of 
equation (26) will correspond to the solution of equation (29) with $\rho(x)$ 
replaced by $\rho^{(L*)}(x)$, where
\be 
\rho^{(L*)}(x)\equiv 
x^{2}(1+cx+\rho_{2}x^{2}+\sum_{k=3}^{\infty}\rho_{k}^{(L)}x^{k})
\;.
\ee
For the observables to which we shall apply the resummation exact NNLO results 
exist and so we have included the exact $\rho_{2}$, rather than 
$\rho_{2}^{(L)}$.

We can define $\rho^{(L*)}$ using the principal value (P.V.) regulated
Borel sum 
results for $D^{(L)}(a)$ of equation (10) obtained in reference \cite 
{charles2}.
\be
D^{(L)}(a)=P.V.\int_{0}^{\infty}\mbox{d}z\,\mbox{e}^{-z/a}B[D^{(L)}](z)
\;,
\ee
where $B[D^{(L)}](z)$ denotes the Borel transform which potentially contains 
poles at $z$=$z_{l}$=$\frac{2l}{b}$ ($l$=$1,2,3,\cdots$) corresponding to 
infra-red renormalons (I${\rm R}_{l}$), and at $z$=$-z_{l}$ corresponding to 
ultra-violet renormalons (U${\rm V}_{l}$). The I${\rm R}_{l}$ singularities are 
intimately connected with the operator product expansion (OPE) for the 
observable in question, and the chosen regulation of the IR singularities 
determines the definition of non-perturbative condensates \cite {grun3}.

In reference \cite {charles2} results have been derived for the $e^{+}e^{-}$ 
Adler 
$D$-function ($\tilde D$) and the polarized Bjorken (or GLS) DIS sum rules 
($\tilde K$). For these Euclidean quantities one can obtain the regulated Borel 
sum of equation (32) as sums of exponential integral functions $Ei(Fz_{l})$ and 
$Ei(-Fz_{l})$, where $F\equiv \frac{1}{a}$. The resulting expressions for 
$\tilde 
D^{(L)}(F)$ and  $\tilde K^{(L)}(F)$ split into UV and IR contributions are 
given in equations (48,49) and equations (52,53) respectively in reference 
\cite {charles2}. 
Results are also obtained for two Minkowski quantities, the $e^{+}e^{-}$ 
$R$-ratio ($\tilde R$) and the analogous $\tau$-decay ratio
($\tilde R_{\tau}$). 
Expressions for $\tilde R^{(L)}(F)$ and $\tilde R_{\tau}^{(L)}(F)$ are given in 
equations (60,64) and equations (69,70) respectively  of reference \cite 
{charles2}, in 
terms of generalized exponential integral functions.

Equation (18) at the leading-$b$ level ($\beta(a)$=$a^{2}$) yields
\be
\rho^{(L)}(x)=(a^{(L)}(x))^{2}\frac{\mbox{d}D^{(L)}(a)}{\mbox{d}a}
\Bigg|_{a=a^{(L)}(x)}\;,
\ee
where $a^{(L)}(x)$ is the inverse function to $D^{(L)}(a)$, i.e. 
$D^{(L)}(a^{(L)}(x))$=$x$, and explicitly
\be 
\rho^{(L)}(x)\equiv x^{2}(1+\sum_{k=2}^{\infty}\rho_{k}^{(L)}x^{k})\;.
\ee
$\rho^{(L)}(x)$ can then be straightforwardly obtained from the corresponding 
$D^{(L)}(F)$ expressions of \cite {charles2} for the various observables. For a 
given $x$ one 
numerically solves
\be
D^{(L)}(F(x))=x\;,
\ee
to obtain $F(x)$, and then from equation (33)
\be
\rho^{(L)}(x)=-\frac{\mbox{d}}{\mbox{d}F}D^{(L)}(F)\Bigg|_{F=F(x)}\;.
\ee
Finally comparing equations (31) and (34) one has
\be
\rho^{(L*)}(x)=\rho^{(L)}(x)+cx^{3}+\rho_{2}^{(NL)}x^{4}\;,
\ee
where $\rho_{2}^{(NL)}\equiv \rho_{2}-\rho_{2}^{(L)}$. This
$\rho^{(L*)}(x)$
can 
be 
inserted in equation (29) and the integral performed numerically. Given a value 
of $\Lambda_{\MS}$ and including the known exact NLO result for 
$d_{1}^{\MS}(\mu=Q)$ one can then solve the transcendental equation (29) for 
$D$=$D^{(L*)}$. Conversely given $D$=$D^{(L*)}$=$D_{data}$, from the 
experimental measurement of the observable, one can solve equation (29) for 
$\Lambda_{\MS}$. By varying $Q$, with $\Lambda_{\MS}^{(N_{f})}$ and 
$d_{1}^{\MS}(\mu=Q)$ evaluated with the number of active quark flavours, 
$N_{f}$, changing across quark thresholds, one can study the $Q$-dependence of 
$D^{(L*)}(Q)$. The resummed result $D^{(L*)}$ can also be compared with $\rm 
N^{n}$LO fixed-order perturbative results. Since $d_{1}$ and $d_{2}$
are exactly 
included in any RS one has $D^{(L*)(1)}$=$D^{(1)}$, $D^{(L*)(2)}$=$D^{(2)}$; 
where $D^{(L*)(n)}$ denotes the truncations of equation (28), and $D^{(1)}$, 
$D^{(2)}$ denote the exact NLO and NNLO results.

\begin{figure}[t]
\begin{center}
\mbox{\epsfig{file=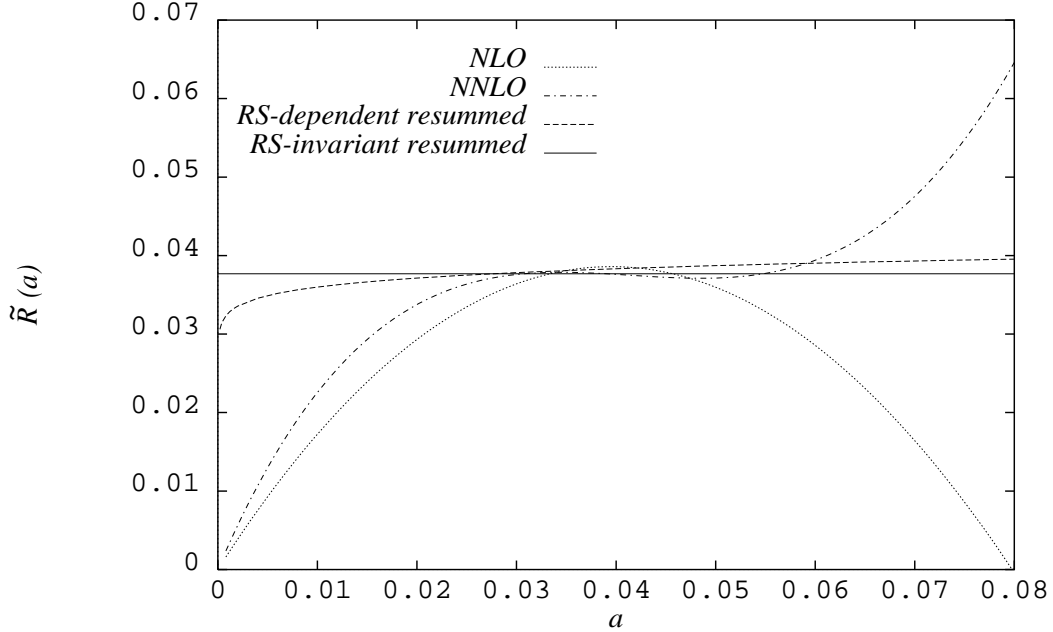,width=5.5in,angle=-90,height=9.6cm}}
\caption{NLO and NNLO fixed-order perturbation theory, the naive RS-dependent 
leading-$b$ 
resummation, and RS-invariant leading-$b$ resummation, for $\tilde R$ at 
$Q=91$ GeV plotted against `$a$'; $c_{2}=0$ has been assumed.}
\end{center}
\end{figure}

In Figure 1(a) we have plotted as the dashed curve the leading-$b$ resummation 
$\tilde R^{(L)}(a)$ versus the coupling `$a$' for the $e^{+}e^{-}$ $R$-ratio 
with $Q$=91 GeV, the t'Hooft scheme corresponding to $B(x)$=$1+cx$, and minimal 
subtraction have been assumed with $\Lambda_{\MS}^{(5)}$=200 MeV. There is a 
monotonic RS-dependence as discussed in reference \cite {charles2}. Noting that 
$\tau$ is 
related to `$a$' using equation (14) one can use `$a$' to label the exact NLO 
and NNLO approximants,  $\tilde R^{(1)}(a)$, $\tilde R^{(2)}(a,c_{2})$. The 
dotted line shows $\tilde R^{(1)}(a)$, and the dashed-dotted line gives $\tilde 
R^{(2)}(a,0)$. We have chosen $c_{2}$=0 to avoid adding an extra axis to the 
plot. The solid line gives the RS-invariant resummation $\tilde R^{(L*)}$. We 
note that the fixed-order results agree best with the $\tilde R^{(L*)}$ 
resummation in the vicinity of the stationary points with respect to variation 
of the RS. This is to be anticipated since the Principal of Minimum Sensitivity 
(PMS) \cite {stev1} choice of RS avoids the inclusion of potentially large UV 
logarithms 
connected with the choice of renormalization scale \cite {reader}. A similar 
statement 
holds for the NLO and NNLO results in the EC scheme \cite {reader}, $\tilde 
R^{(1)}(EC)$ and $\tilde R^{(2)}(EC)$, corresponding to solutions of 
equation (29) with $\rho^{(L*)}$ in equation (31) truncated. These are 
numerically very close to the PMS approximants. The `optimized' PMS/EC 
fixed-order NNLO approximant is thus seen to be very close to to the 
RS-invariant resummed result for the $R$-ratio at LEP energy, indicating that 
the approximated effect of including $\rm N^{3}$LO and higher corrections is 
small, and thus suggesting that one can in principle accurately determine 
$\Lambda_{\MS}^{(5)}$ given ideal data.

\begin{figure}[t]
\begin{center}
\mbox{\epsfig{file=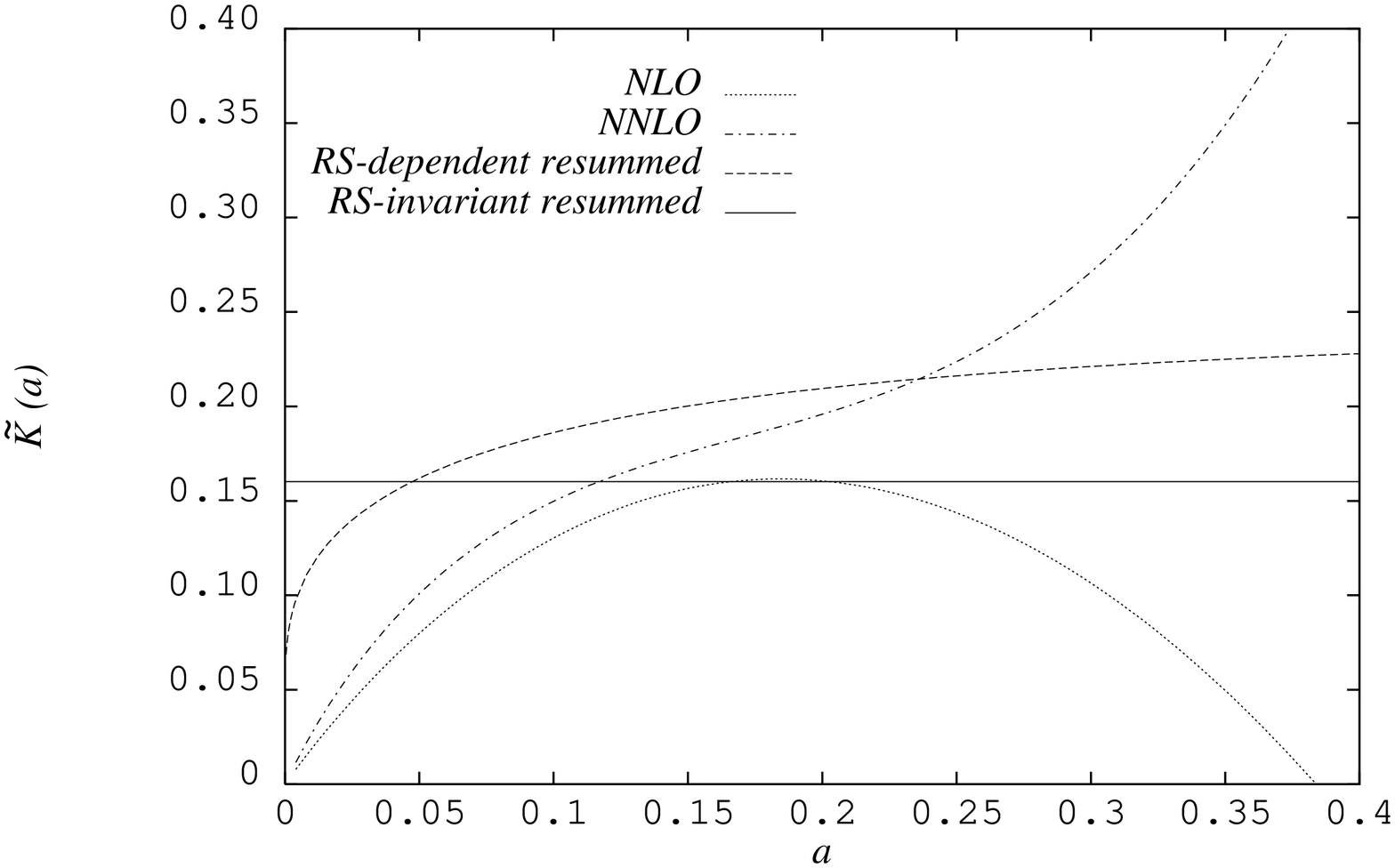,width=5.5in,angle=-90,height=9.6cm}}
\caption{As for Figure 1(a) but for $\tilde K$ at $Q=1.5$ GeV.}
\end{center}
\end{figure}
\begin{figure}[t]
\begin{center}
\mbox{\epsfig{file=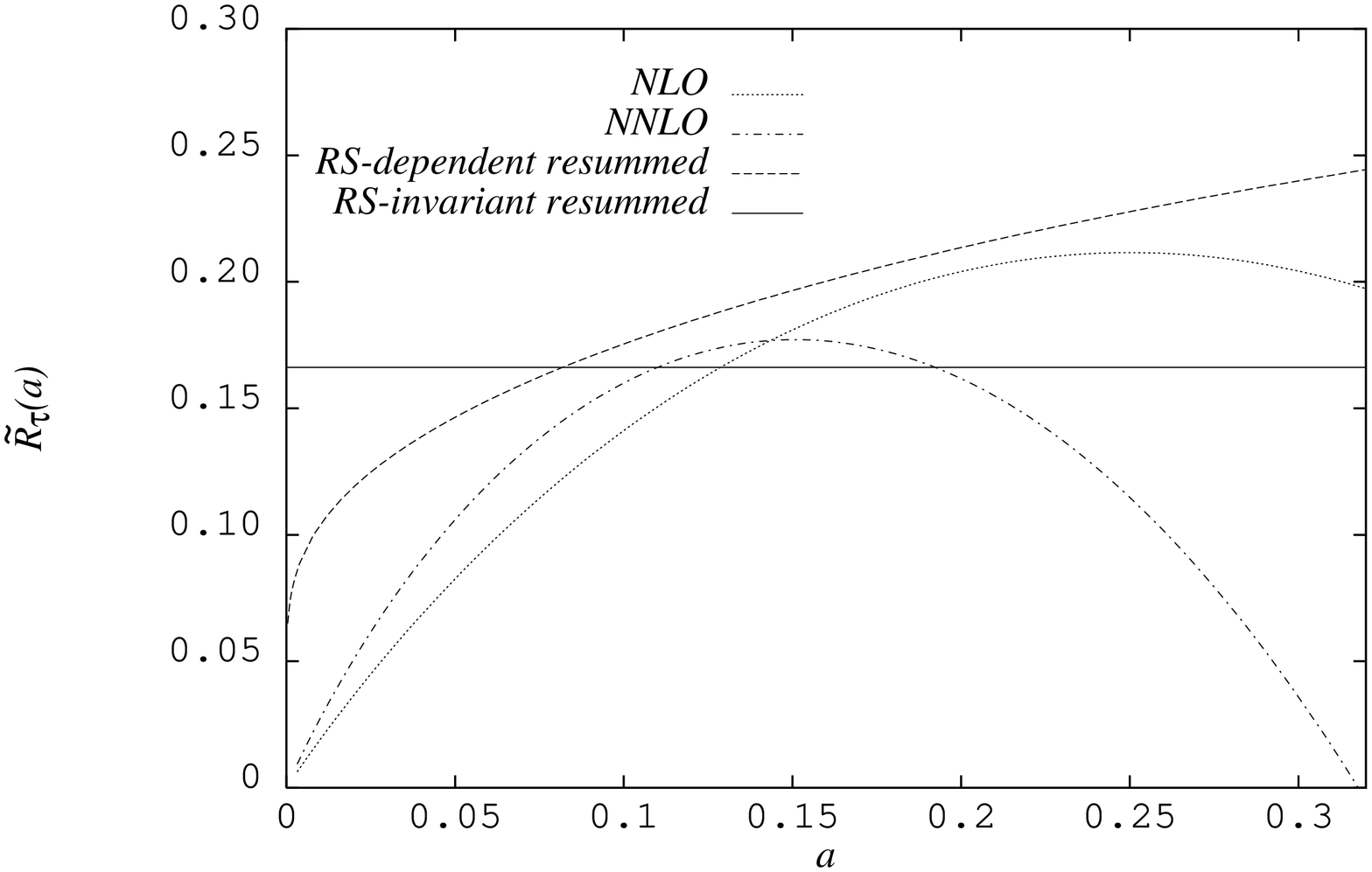,width=5.5in,angle=-90,height=9.6cm}}
\caption{As for Figure 1(a) but for $\tilde R_{\tau}$ at $Q=1.78$ GeV.}
\end{center}
\end{figure}

In Figures 1(b) and 1(c) the analogous plots for the DIS sum rule $\tilde 
K$ at $Q$=1.5 GeV, and for the $\tau$-decay ratio $\tilde R_{\tau}$ 
($Q$=$m_{\tau}$=1.78 GeV) have been given. $\Lambda_{\MS}^{(3)}$=320 MeV has 
been 
assumed.

In contrast to Figure 1(a) the differences between the fixed-order results and 
the RS-invariant resummations are clearly much larger. Thus at these lower 
values 
of $Q$ the significance of $\rm N^{3}$LO and higher effects is apparently much 
greater, and the reliability with which $\Lambda_{\MS}^{(3)}$ can be determined 
correspondingly less. We shall quantify this more carefully in just a moment.

In Figure 2(a) we plot, for the $e^{+}e^{-}$ $R$-ratio at $Q$=91 GeV, the 
fixed-order perturbative results $\tilde R^{(L*)(n)}(EC)$ (equation (28)) for 
$n$=2 (NNLO) and higher orders (crosses) compared with the RS-invariant 
resummed result $\tilde R^{(L*)}$ (dashed line). $\Lambda_{\MS}^{(5)}$=200 MeV 
has 
again been assumed. We could of course have chosen to plot the fixed-order 
approximants in any RS, for instance $\MS$ with $\mu$=$Q$, but as discussed in 
connection with Figure 1(a), we expect the `optimized', EC or PMS, choice of RS 
to approach the resummed result more rapidly. We stress that the fixed-order 
$\tilde R^{(L*)(n)}(EC)$ approximants correspond to the solutions of equation 
(29) with $\rho(x)$ replaced by the truncation of $\rho^{(L*)}(x)$ in equation 
(31). Since $\rho_{2}$ is included exactly $\tilde R^{(L*)(2)}(EC)$=$\tilde 
R^{(2)}(EC)$.

\setcounter{figure}{0}
\renewcommand{\thefigure}{2(\alph{figure})}

\begin{figure}[p]
\begin{center}
\mbox{\epsfig{file=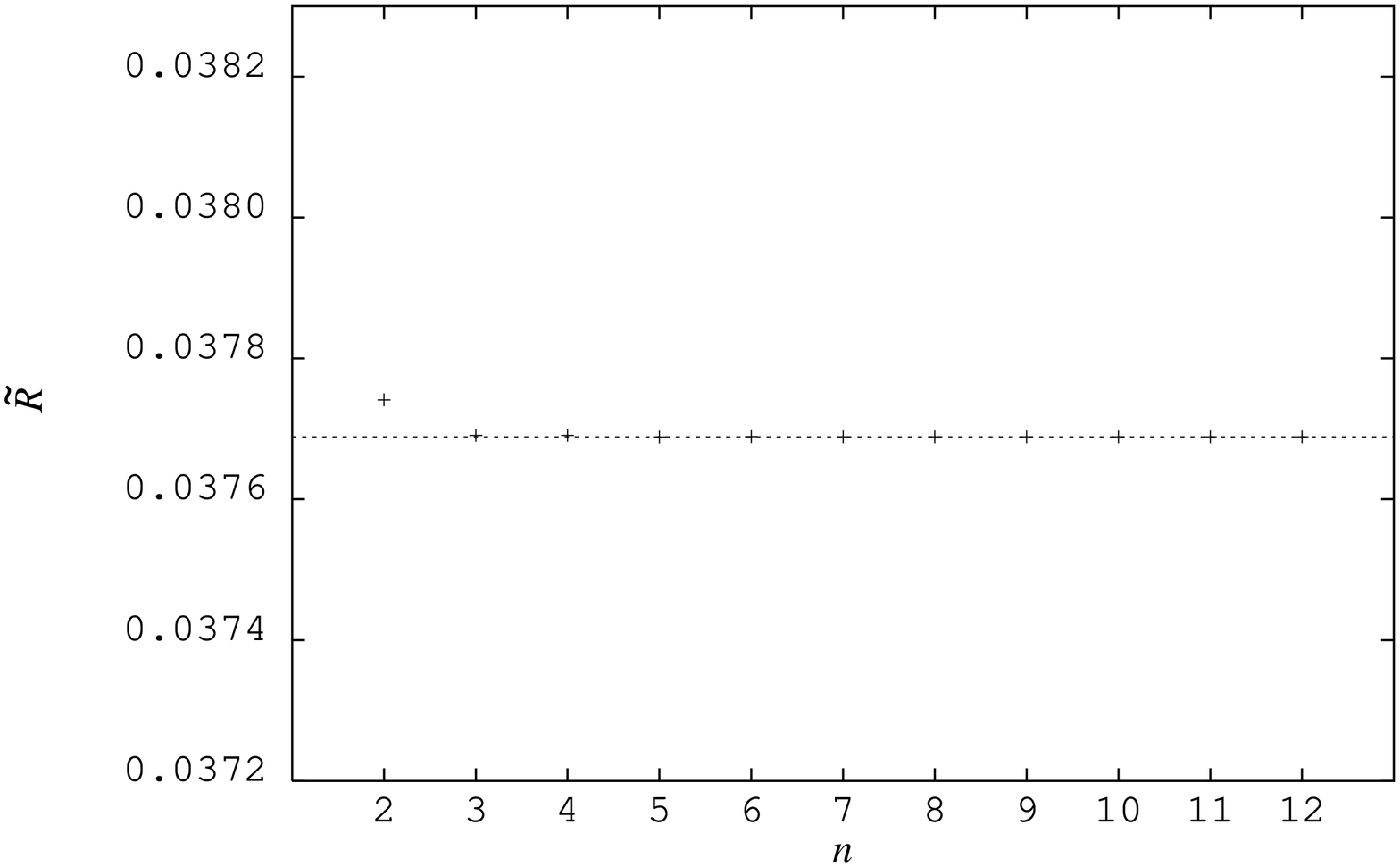,width=5.5in,angle=-90,height=9.5cm}}
\caption{Comparison of fixed-order EC perturbation theory (crosses) with the 
RS-invariant 
resummation (dashed line) for $\tilde R$ at $Q=91$ GeV.}
\end{center}
\end{figure}
\begin{figure}[p]
\begin{center}
\mbox{\epsfig{file=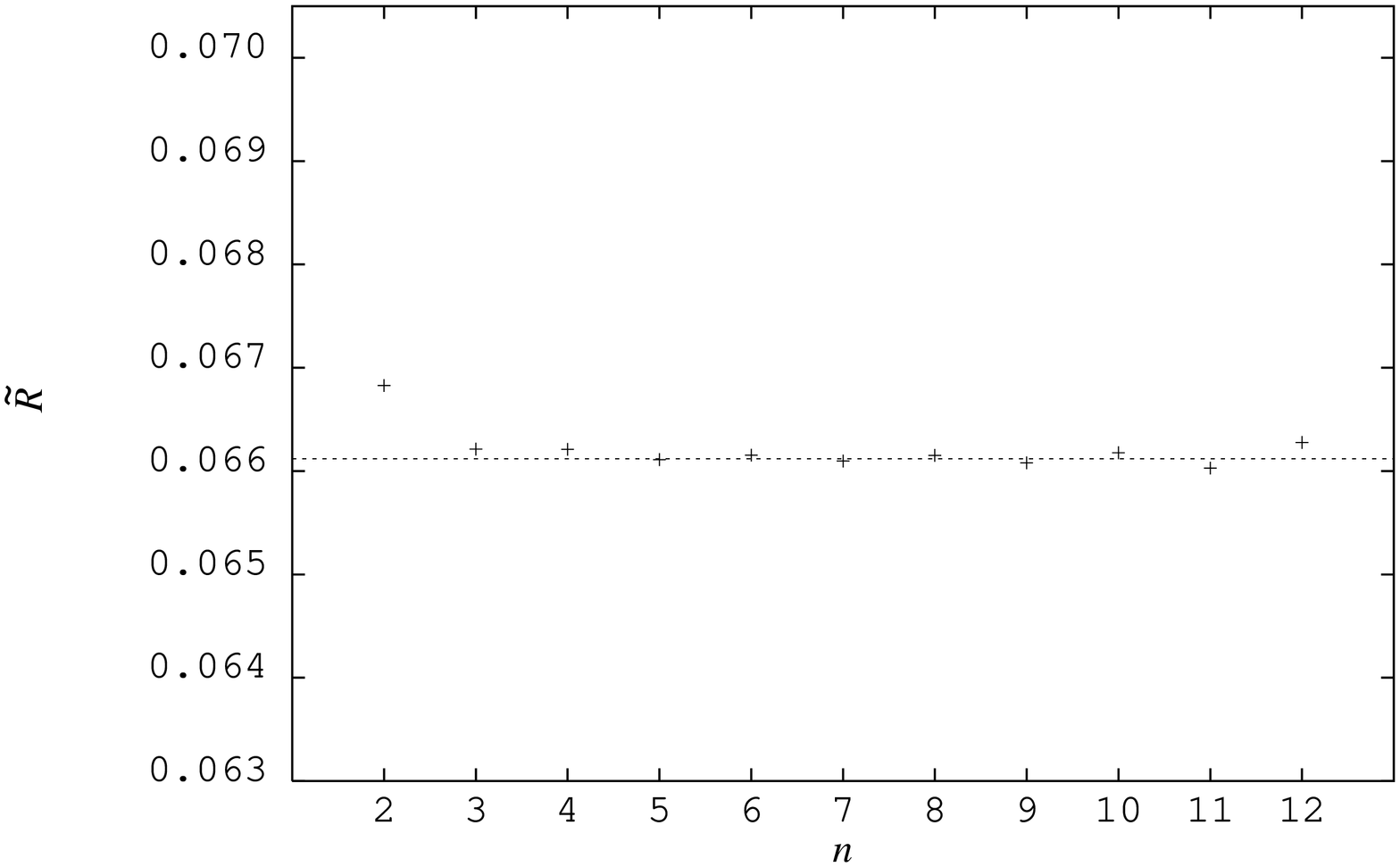,width=5.5in,angle=-90,height=9.5cm}}
\caption{As for Figure 2(a) except at $Q=5$ GeV.}
\end{center}
\end{figure}
\begin{figure}[t]
\begin{center}
\mbox{\epsfig{file=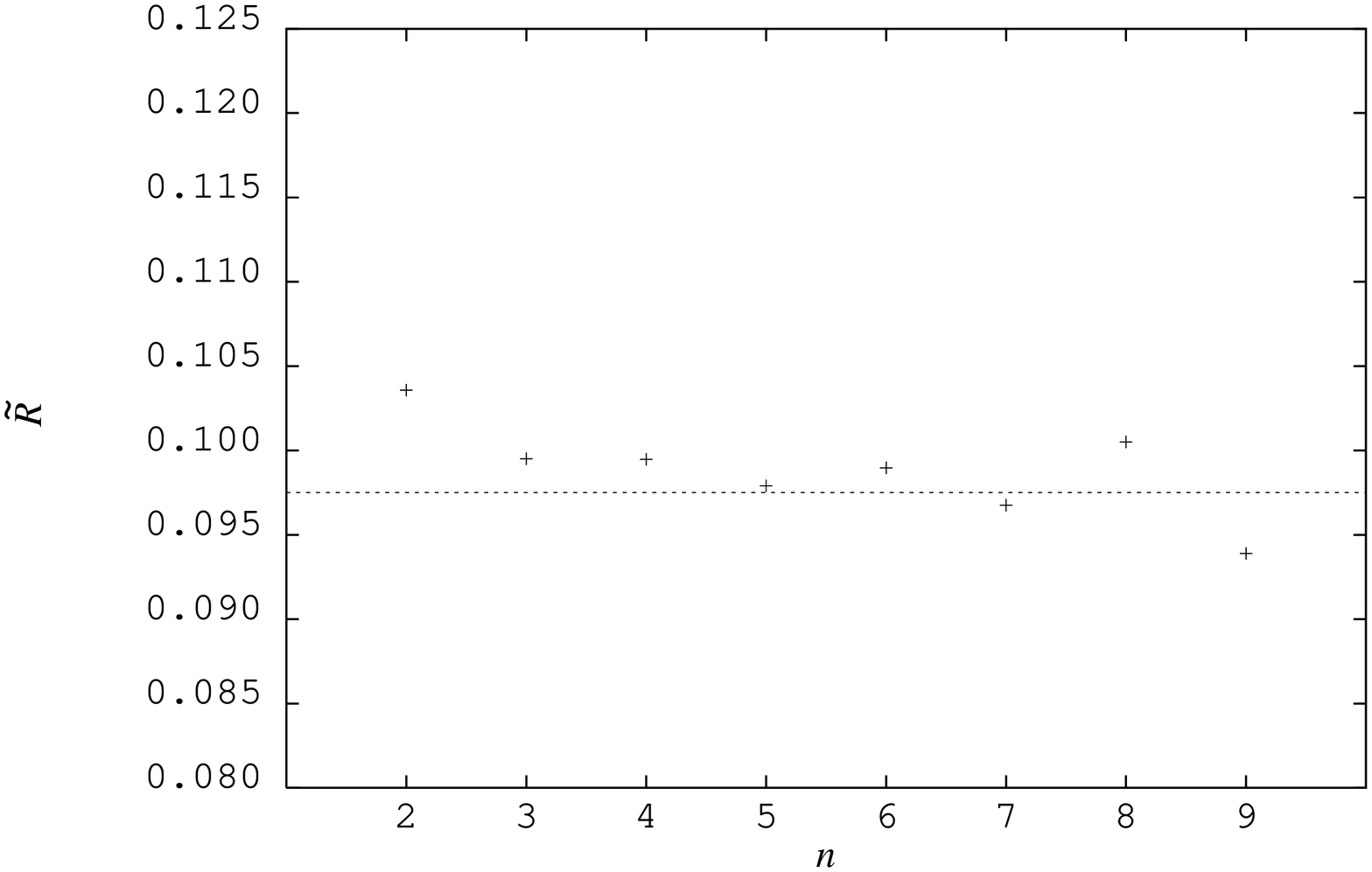,width=5.5in,angle=-90,height=9.5cm}}
\caption{As for Figure 2(a) except at $Q=1.5$ GeV.}
\end{center}
\end{figure}

As can be seen from Figure 2(a) the $\rm N^{3}$LO and higher
fixed-order results 
are indistinguishable from the resummed result with the chosen vertical scale, 
and there is only a small shift between the NNLO and resummed results.
Evidently 
fixed-order perturbation theory in the EC scheme seems to be working very well 
for the $R$-ratio at LEP/SLD energies.

In Figure 2(b) we show a similar plot for $\tilde R$ at $Q$=5 GeV. 
$\Lambda_{\MS}^{(4)}$=279 MeV has been assumed. Clearly the approach to the 
resummed result is somewhat less rapid. The slight oscillation of successive 
fixed-order approximants above and below the resummed result is
explained by the 
dominance of the U$\rm V_{1}$ singularity at $z$=$-\frac{2}{b}$ in the Borel 
plane, which is the closest to the origin for the $R$-ratio.
This singularity is 
responsible for alternating sign factorial growth of the perturbative 
coefficients. Beyond order $n$=12 the amplitude of the oscillations increases 
dramatically, and the fixed-order approximants diverge increasingly from the 
resummation. This is precisely what one would expect to see on comparing the 
Borel sum of an alternating sign asymptotic series with its
truncations.
We note 
that a similar oscillating behaviour with eventual wild oscillations setting in 
would also have been apparent in Figure 2(a) had we used a finer
vertical scale. 
>From the large-order behaviour one would not expect the wild
oscillations to set 
in until $n>50$.

Figure 2(c) finally shows the corrsponding plot for $\tilde R$ at $Q$=1.5 GeV, 
with $\Lambda_{\MS}^{(3)}$ as above. The approach to the resummed result is 
still 
slower, and the oscillations have only just become established when they 
increase wildly beyond $n$=9. Nonetheless even at this low energy fixed-order 
perturbation theory is approximating the resummed results, albeit much less 
well.

\setcounter{figure}{0}
\renewcommand{\thefigure}{3(\alph{figure})}

\begin{figure}[p]
\begin{center}
\mbox{\epsfig{file=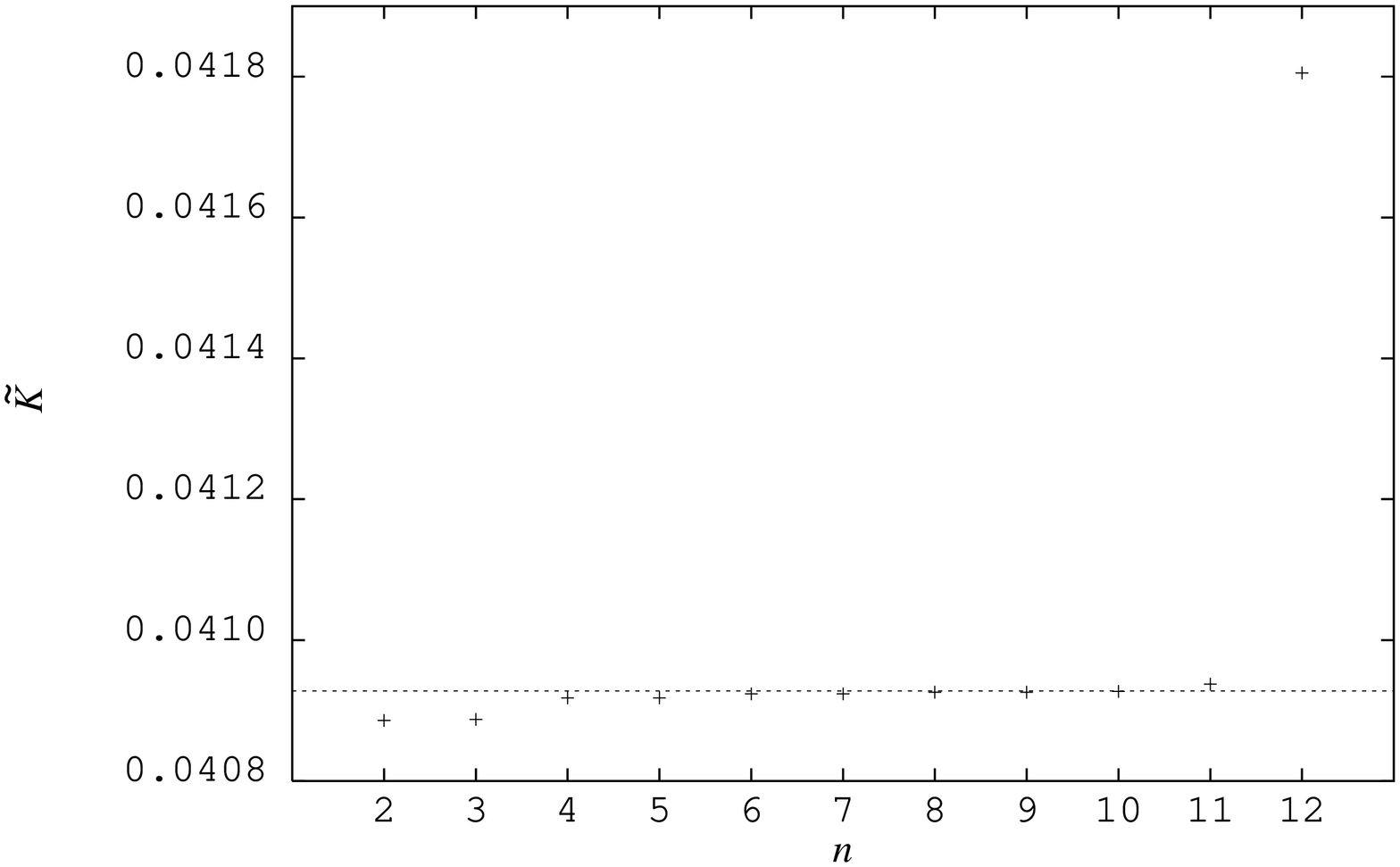,width=5.5in,angle=-90,height=9.5cm}}
\caption{Comparison of fixed-order EC perturbation theory (crosses) with the 
RS-invariant 
resummation (dashed line) for $\tilde K$ at $Q=91$ GeV.}
\end{center}
\end{figure}
\begin{figure}[p]
\begin{center}
\mbox{\epsfig{file=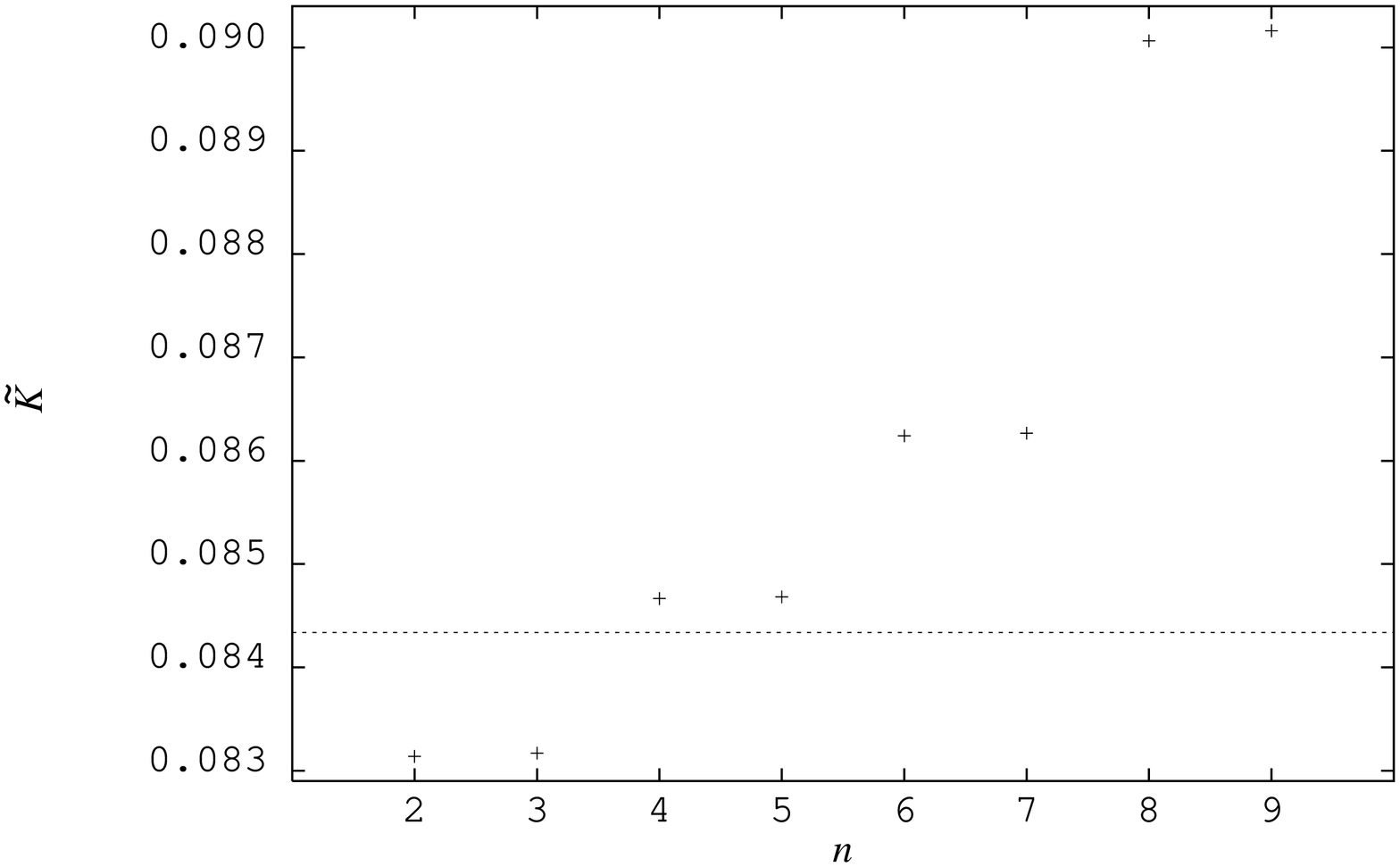,width=5.5in,angle=-90,height=9.5cm}}
\caption{As for Figure 3(a) except at $Q=5$ GeV.}
\end{center}
\end{figure}
\begin{figure}[t]
\begin{center}
\mbox{\epsfig{file=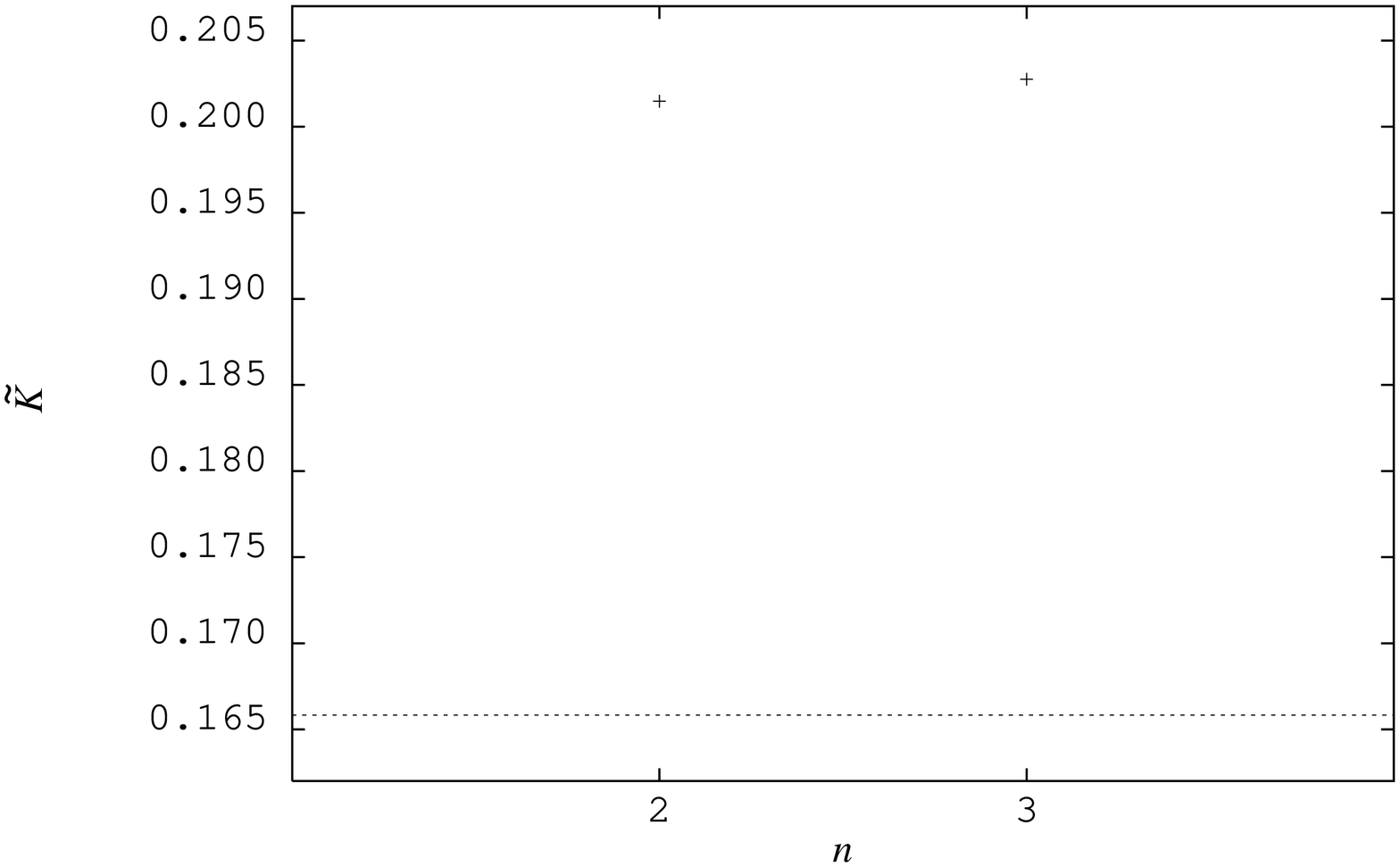,width=5.5in,angle=-90,height=9.5cm}}
\caption{As for Figure 3(a) except at $Q=1.5$ GeV.}
\end{center}
\end{figure}

This reasonable performance of fixed-order perturbation theory
for $\tilde R$ is 
to be contrasted with the situation for the DIS sum rules $\tilde K$. Figures 
3(a), 3(b), and 3(c) are plotted at the same values of $Q$ as the corresponding 
Figures 2 for $\tilde R$, and with the same vertical scales to enable direct 
comparisons. In Figure 3(a) at $Q$=91 GeV we see a much slower approach to the 
resummed result. The fixed-order EC approximations then track the resummed 
result between sixth and tenth order and then for $n$=12 there is a dramatic 
breakdown. The Borel plane singularities nearest the origin for $\tilde K$ are 
now \cite {charles2} I$\rm R_{1}$ at $z$=$\frac{2}{b}$, and U$\rm V_{1}$ at 
$z$=$-\frac{2}{b}$. It is the presence of the I$\rm R_{1}$ singularity which 
leads to fixed-sign factorial growth of the perturbative coefficients and a 
consequent deterioration in the performance of fixed-order perturbation theory. 
The relative deterioration compared to the $R$-ratio can be seen even more 
clearly in Figure 3(b) at $Q$=5 GeV. There is a monotonic increase
in successive 
orders with no tendency to track the resummed result. The stair-like pattern, 
with neighbouring odd and even orders roughly similar in low orders, follows 
from a partial cancellation between the fixed-sign (I$\rm R_{1}$) and 
alternating-sign (U$\rm V_{1}$) behaviours.

In Figure 3(c) we see that at $Q$=1.5 GeV fixed-order EC perturbation theory is 
a 
poor approximation to the resummed result for the DIS sum rules. Only $n$=2 and 
$n$=3 are shown since for $n\geq$4 fixed-order perturbation theory is not 
defined in the EC scheme. If $\rho(x)$ has a zero at $x$=$x^{*}$ (where 
$x^{*}>0$ is the closest zero to the origin) then equation (29) has a solution 
$D=D^{*}$, with $D^{*}<x^{*}$. This will be the case if the expansion 
coefficients of $\rho(x)$ have alternating factorial behaviour, at least in 
either odd or even orders. If, however, the coefficients have fixed-sign 
factorial growth, as is the case for the DIS sum rules, then
$\rho(x)$ will have 
no positive zeros. In this case equation (29) may fail to have a solution, the 
condition for this being that in the limit as $D\rightarrow+\infty$ the 
right-hand side of equation (29) is negative.

In Figure 4 we show the analogous plot for the hadronic $\tau$-decay ratio, 
$R_{\tau}$. Here evidently $Q=m_{\tau}=1.78$ GeV, and the same 
$\Lambda_{\MS}^{(3)}$ as above has been assumed. Fixed-order EC perturbation 
theory 
is seen to be working reasonably well with oscillating behaviour around the 
resummed result which becomes wild for $n\geq5$. Notice, however, that the 
performance is much worse than that of $\tilde R$ at the comparable $Q=1.5$ GeV 
in Figure 2(c).

\setcounter{figure}{0}
\renewcommand{\thefigure}{4}

\begin{figure}[t]
\begin{center}
\mbox{\epsfig{file=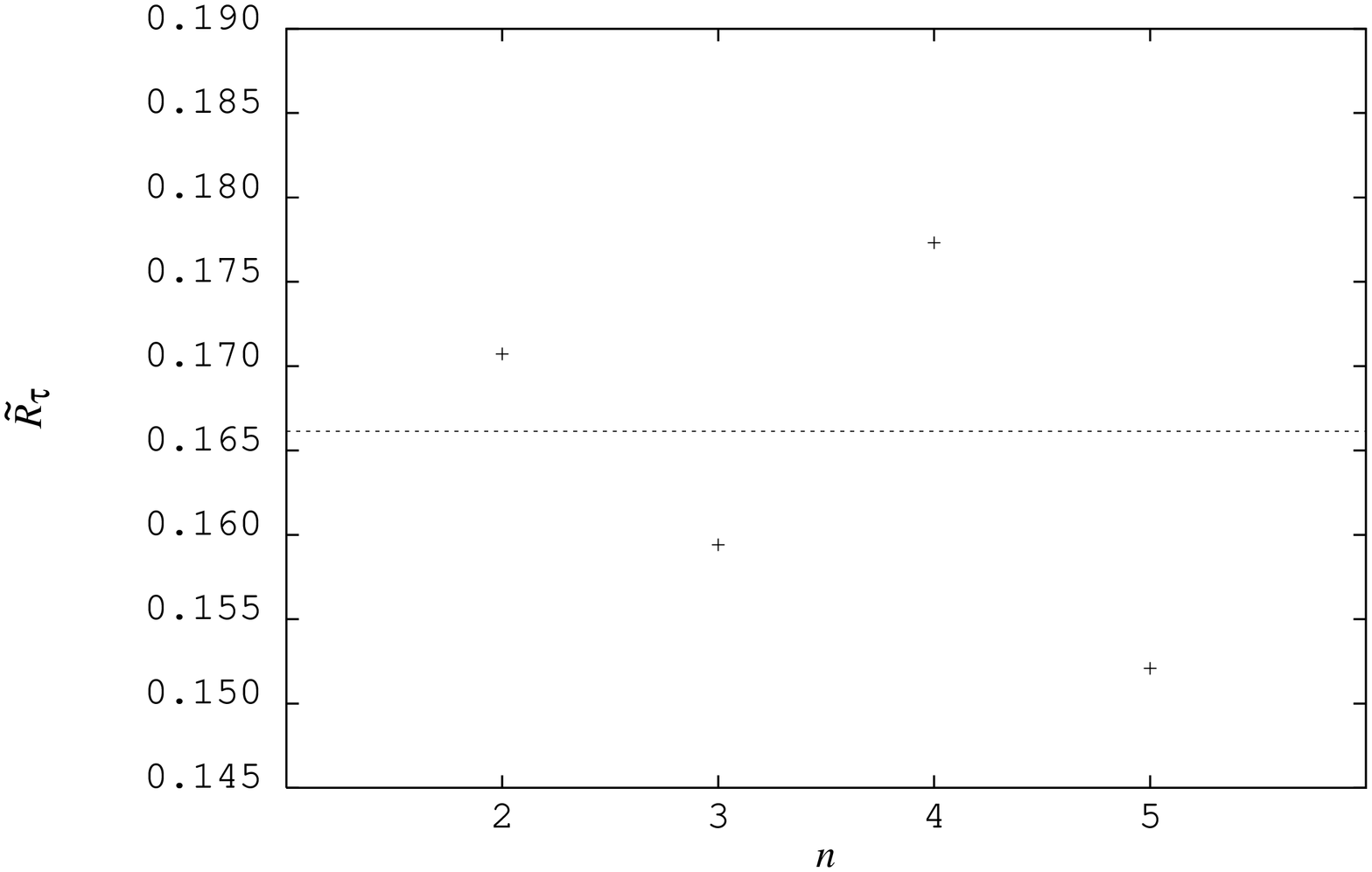,width=5.5in,angle=-90,height=9.6cm}}
\caption{Comparison of fixed-order EC perturbation theory (crosses) with the 
RS-invariant 
resummation (dashed line) for $\tilde R_{\tau}$.}
\end{center}
\end{figure}

\setcounter{figure}{0}
\renewcommand{\thefigure}{5(\alph{figure})}

\begin{figure}[p]
\begin{center}
\mbox{\epsfig{file=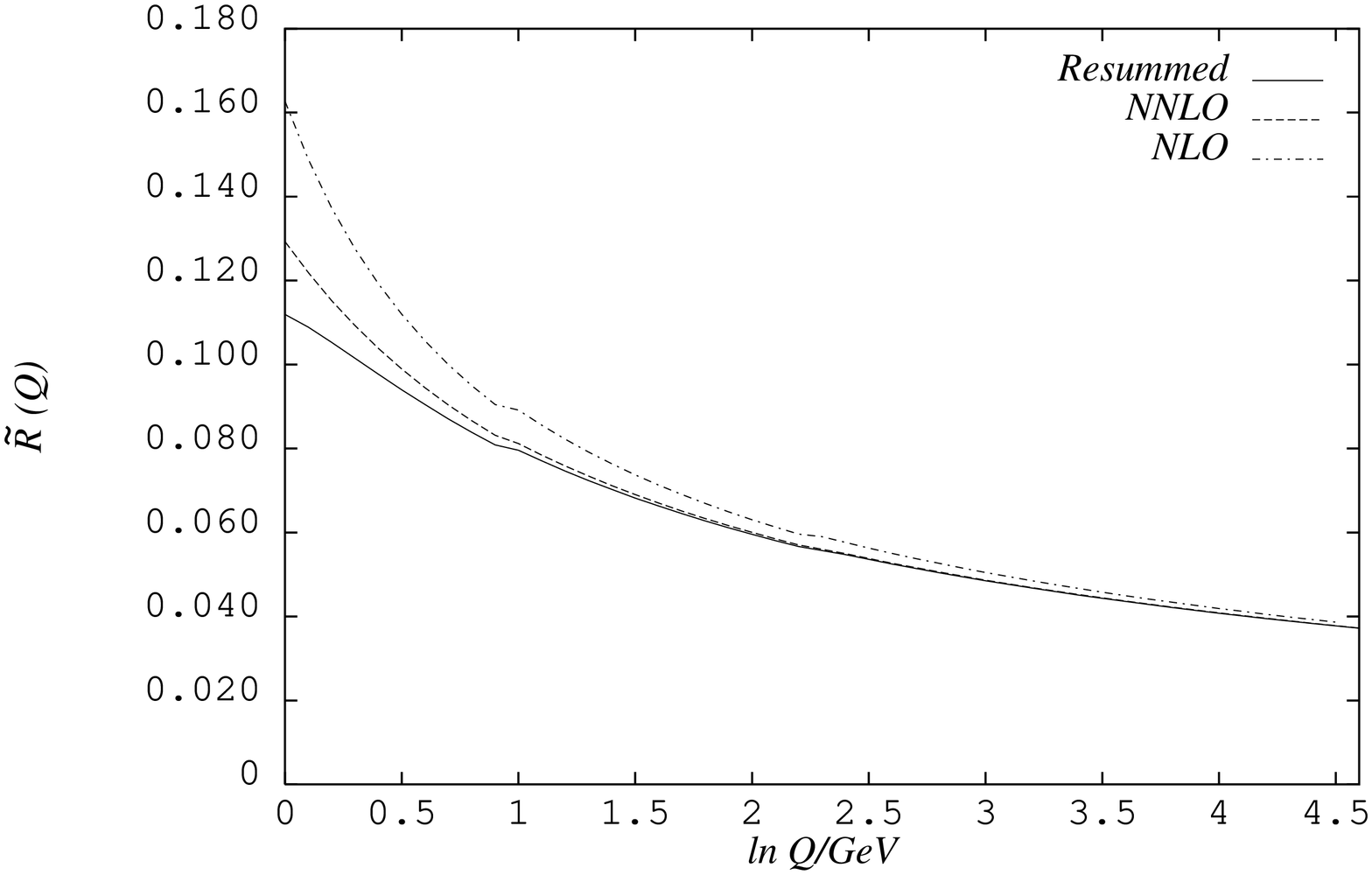,width=5.5in,angle=-90,height=9.5cm}}
\caption{NLO, NNLO fixed-order results in the EC scheme, and RS-invariant 
resummation, for 
$\tilde R$ plotted versus $\ln Q$/GeV over the range $1\leq Q\leq 91$ GeV.}
\end{center}
\end{figure}
\begin{figure}[p]
\begin{center}
\mbox{\epsfig{file=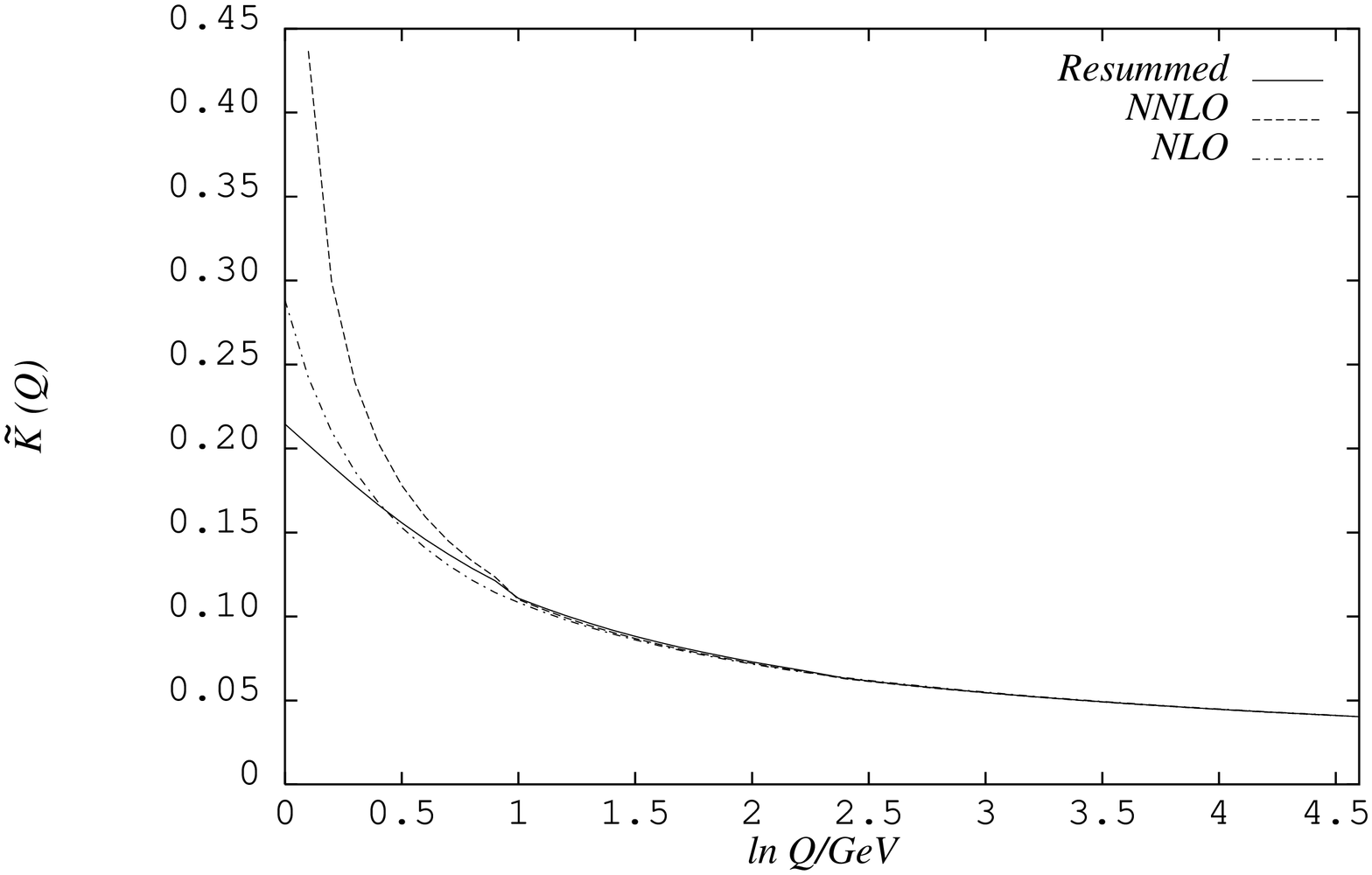,width=5.5in,angle=-90,height=9.5cm}}
\caption{As for Figure 5(a) except for $\tilde K$.}
\end{center}
\end{figure}

We can summarize the behaviours exhibited in the foregoing figures by plotting 
the energy dependence of $\tilde R$ and $\tilde K$. Figure 5(a) shows $\tilde 
R^{(L*)}$ (solid line), $\tilde R^{(1)}(EC)$ (dotted line), and $\tilde 
R^{(2)}(EC)$ (dashed-dotted line), plotted versus \( \ln Q/GeV \)over a range 
equivalent to $Q=1-91$ GeV. Flavour thresholds in $Q$ at $m_{b}=4.5$ GeV, 
$m_{c}=1.25$ GeV, have been assumed and values of $\Lambda_{\MS}^{(N_{f})}$ 
chosen 
as above. Figure 5(b) gives a similar plot for $\tilde K$. We note the 
reasonably satisfactory behaviour for $\tilde R$, in particular at all energies 
the NNLO EC approximation is closer to the resummed result than the NLO, as one 
would hope. In contrast for $\tilde K$ below $Q\sim3$ GeV the 
NLO becomes closer 
than NNLO to the resummed result, making the use of fixed-order perturbation 
theory dubious. Notice in addition that the vertical scale in Figure 5(b) is 
much coarser than that in Figure 5(a).

We would finally like to use the RS-invariant all-orders resummation to assess 
the likely accuracy to which $\Lambda_{\MS}^{(N_{f})}$ can be determined for 
various observables at various energies.

\begin{table}[t]
\begin{center}
\begin{tabular}{|c|c|c|c|c|c|c|}\hline
{\raisebox{-12pt}{Observable}} & {\raisebox{-12pt}{Energy $Q$/GeV}} & 
{\raisebox{-12pt}{$N_{f}$}} & {\raisebox{-12pt}{Data}} &
\multicolumn{3}{|c|}{{\raisebox{-12pt}{$\Lambda_{\MS}^{(N_{f})}$/MeV fitted 
to experiment}}}\\ 
& & & & \multicolumn{3}{|c|}{ }\\ \cline{5-7}
 &  &  &  & {\raisebox{-12pt}{NLO}} & {\raisebox{-12pt}{NNLO}}
 & {\raisebox{-12pt}{Resummed}} \\
 & & & & & & \\ \hline
 & & & & & & \\
{\raisebox{+9pt}{$\tilde R$}} & {\raisebox{+9pt}{91}} & {\raisebox{+9pt}{5}} & 
{\raisebox{+9pt}{$0.040\pm0.004$}} & {\raisebox{+9pt}{$252^{+190}_{-126}$}} & 
{\raisebox{+9pt}{$293^{+228}_{-149}$}} & {\raisebox{+9pt}{$296^{+232}_{-150}$}}
\\ \cline{2-7}
 & & & & & & \\ 
 & {\raisebox{+9pt}{9}} & {\raisebox{+9pt}{5}} & 
{\raisebox{+9pt}{$0.073\pm0.024$}}
 & {\raisebox{+9pt}{$399^{+478}_{-322}$}} & 
{\raisebox{+9pt}{$516^{+735}_{-423}$}} 
 & {\raisebox{+9pt}{$537^{+823}_{-443}$}}\\ \hline
 & & & & & & \\
{\raisebox{+9pt}{$\tilde R_{\tau}$}} & {\raisebox{+9pt}{1.78}} & 
{\raisebox{+9pt}{3}} & {\raisebox{+9pt}{$0.205\pm0.006$}} & 
{\raisebox{+9pt}{$319^{+7}_{-7}$}}
& {\raisebox{+9pt}{$387^{+11}_{-11}$}} & {\raisebox{+9pt}{$404^{+12}_{-12}$}}
\\ \hline
& & & & & & \\
{\raisebox{+9pt}{$\tilde K$}} & {\raisebox{+9pt}{2.24}}
& {\raisebox{+9pt}{3}} & 
{\raisebox{+9pt}{$0.154\pm0.075$}} & {\raisebox{+9pt}{$437^{+187}_{-299}$}} & 
{\raisebox{+9pt}{$379^{+138}_{-252}$}} & {\raisebox{+9pt}{$426^{+386}_{-303}$}}
\\ \hline
\end{tabular}
\caption[t]{Values of $\Lambda_{\MS}^{(N_{f})}$ adjusted to fit the predictions
of NLO, NNLO fixed-order results in the EC scheme, and the RS-invariant 
resummation
to the experimental data for $\tilde R$, $\tilde R_{\tau}$ 
and $\tilde K$.}
\end{center}
\end{table}

In Table 1 we have given the $\Lambda_{\MS}^{(N_{f})}$ values obtained by 
fitting 
the NLO, NNLO EC fixed-order perturbative and the RS-invariant resummed results 
to the central values of the data for $\tilde R$($Q$=91 GeV) \cite {lep}, 
$\tilde 
R$($Q$=9 GeV) \cite {marshall}, $\tilde R_{\tau}$, and $\tilde K$($Q^{2}$=5 
Ge$\rm V^{2}$). The value for $\tilde R_{\tau}$ is that obtained in \cite 
{bert3} by averaging those obtained from the leptonic branching ratio \cite 
{rahal} and $\tau$-lifetime measurements \cite {wasser}; we have corrected for 
the 
small estimated power corrections \cite {bert2}.

For $\tilde K$($Q^{2}=5\rm GeV^{2}$) we have taken the GLS sum rule result of 
the CCFR collaboration \cite {ccfr} corrected by subtracting off the $Q^{-2}$ 
higher-twist corrections suggested by reference \cite{braun}, so that
\be
\tilde K(Q)=\tilde K_{CCFR}(Q)+\frac{(0.27\pm0.14)}{Q^{2}}\;.
\ee
The errors have been combined in quadrature.

The results in Table 1 are encouraging in that they indicate rather small 
differences between the NNLO EC and resummed fits. From these differences one 
would estimate that one could determine $\Lambda_{\MS}^{(5)}$ to an accuracy of 
$\sim\pm$3 MeV at LEP/SLD energies given ideal data for $\tilde R$, 
corresponding to determining $\alpha_{s}(M_{Z})$ to three significant figures. 
Needless to say even with ideal data undetermined finite quark mass effects 
would in fact introduce far larger uncertainties.

At $Q=9$ GeV $\Lambda_{\MS}^{(5)}$ would apparently be determined to
an accuracy 
of $\sim\pm20$ MeV. The data for $\tilde R$($Q=91$ GeV) imply NNLO 
$\alpha_{s}(M_{Z})$ ($\MS$) values of $\alpha_{s}(M_{Z})$=$0.122\pm0.012$; NNLO 
EC 
and resummed are the same to the quoted number of significant figures.

For $\tilde R_{\tau}$ a comparison of the NNLO EC and resummed fits would 
suggest that $\Lambda_{\MS}^{(3)}$ could be determined with a precision of 
$\sim\pm15$ MeV. One finds $\alpha_{s}(m_{\tau})$=$0.320\pm0.005$ (NNLO EC) and 
$\alpha_{s}(m_{\tau})$=$0.328\pm0.005$ (resummed). Evolving up
from $N_{f}=3$ to 
$N_{f}=5$ assuming the flavour thresholds noted above yields 
$\Lambda_{\MS}^{(5)}$=$253^{+9}_{-9}$ MeV (NNLO EC) and  
$\Lambda_{\MS}^{(5)}$=$267^{+10}_{-10}$ MeV (resummed), corresponding 
to $\alpha_{s}(M_{Z})$=$0.119\pm0.001$ (NNLO EC) and 
$\alpha_{s}(M_{Z})$=$0.120\pm0.001$ (resummed). A conservative estimate of the 
theoretical uncertainty is then 
$\delta\alpha_{s}(M_{Z})$=$0.001$.

If taken seriously the above estimate of the accuracy with which 
$\Lambda_{\MS}^{(5)}$ ($\alpha_{s}(M_{Z})$) can be determined from $R_{\tau}$ 
measurements is very reassuring, and clearly indicates that this is indeed 
potentially the most reliable determination. The uncertainty is 
somewhat smaller 
than has been assumed based on more naive estimates of the size of the 
$O(a^{4})$ perturbative coefficient \cite {pich}. It is much smaller than 
$\delta\alpha_{s}(m_{\tau})=0.05$ 
inferred 
by Neubert in reference \cite {bert2} based on comparison of the exact 
$O(\alpha_{s}^{3})$ NNLO perturbative result in the $\MS$ scheme with 
$\mu$=$m_{\tau}$, and a straightforward resummation of the leading-$b$ terms, 
which is essentially our $\tilde R_{\tau}^{(L)}$ (c.f. equation (10)), with 
$a$=$\frac{\alpha_{s}(m_{\tau})}{\pi}$. As can be seen from Figure 1(c) the 
dashed curve $\tilde R_{\tau}^{(L)}(a)$ lies above the RS-invariant resummation 
(solid line) for $a$=$\frac{\alpha_{s}(m_{\tau})}{\pi}\simeq0.12$, and there is 
a strong `$a$' dependence in this region. The NNLO EC result, and indeed the 
$\MS$ $\mu=m_{\tau}$ NNLO result, are both much closer to the RS-invariant 
resummation. The implication then is that the rather large difference between 
the exact fixed-order and naive resummed leading-$b$ results found in \cite 
{bert2} is 
just a reflection of the inadequacy of the naive resummation, which was our 
original motivation for improving it.

We finally turn to the GLS sum rule results in Table 1. Whilst the 
$\Lambda_{\MS}^{(3)}$ values for NLO, NNLO, and resummed are in reasonable 
agreement 
with that obtained for $\tilde R_{\tau}$, we note once again the worrying 
feature that the NLO result is closer to the resummed than the NNLO. We have 
also had to assume and correct for sizeable power corrections, based on the 
modelled 
suggestion of reference \cite {braun}. Also note the very large errors which 
reflect the 
difficulty in reconstructing the sum rule by combining data from various DIS 
experiments \cite {ccfr}. Clearly $\tilde K$ will not be competitive with 
$\tilde 
R_{\tau}$ as a way of determining $\Lambda_{\MS}$.

\section{Discussion}

Before giving a summary of our main conclusions we would like to
discuss several 
ways in which we could improve or extend the RS-invariant resummations, and 
mention some technical issues related to them.

The first concerns the analytical continuation between the Euclidean Adler 
$D$-function and the Minkowski quantities $\tilde R$ and $\tilde
R_{\tau}$.
This 
will imply definite relations between the corresponding RS-invariants 
$\rho_{k}^{D}$, $\rho_{k}^{R}$, $\rho_{k}^{R_{\tau}}$. For instance for $\tilde 
R$ 
one 
has \cite {racz1}
\begin{eqnarray}
\rho_{2}^{R}&=&\rho_{2}^{D}-\frac{1}{12}b^{2}\pi^{2}\nonumber\\
\rho_{3}^{R}&=&\rho_{3}^{D}-\frac{5}{12}cb^{2}\pi^{2}\\
\rho_{4}^{R}&=&\rho_{4}^{D}-\frac{1}{12}(8\rho_{2}^{D}+7c^{2})b^{2}\pi^{2}+
\frac{1}{360}b^{4}\pi^{4}\nonumber\\
\vdots & & \vdots \nonumber 
\end{eqnarray}
The procedure we have used to construct $\tilde R^{(L*)}$ involves 
resumming the 
effective charge beta-function using the exact $\rho_{2}^{R}$ and the 
leading-$b$ approximations to $\rho_{k}^{R}$, $k>2$. This means the 
$-\frac{5}{12}cb^{2}\pi^{2}$ analytical continuation term in $\rho_{3}^{R}$, or 
the $-\frac{7}{12}c^{2}b^{2}\pi^{2}$ in $\rho_{4}^{R}$, have been omitted since 
they are sub-leading in $b$. Since $\rho_{2}^{D}$ is known exactly 
we could also 
improve the  resummation by using the exact $\rho_{2}^{D}$ in evaluating the 
$-\frac{2}{3}\rho_{2}^{D}b^{2}\pi^{2}$ term in $\rho_{4}^{R}$. One could 
envisage an improved resummation $\rho_{R}^{(L**)}(x)$ incorporating 
these extra 
terms.
\be
\rho_{R}^{(L**)}(x)=\rho_{R}^{(L*)}(x)+\tilde \rho_{R}(x)\;,
\ee
where the extra terms to $O(x^{7})$ are explicitly
\be
\tilde 
\rho_{R}(x)=-\frac{5}{12}cb^{2}\pi^{2}x^{5}-(\frac{2}{3}\rho_{2}^{D(NL)}+
\frac{7}{12}c^{2})b^{2}\pi^{2}x^{6}+O(x^{7})+\cdots\;.
\ee

This resummation to all-orders can be accomplished in principle by representing 
$\tilde R$ as a contour integral involving $\tilde D$ \cite {racz1}. Using 
$\tilde 
D^{(L*)}$ in the integrand would formally produce $\tilde R^{(L**)}$ 
corresponding to the above effective charge beta-function $\rho_{R}^{(L**)}$, 
but one would have to evaluate $\tilde D^{(L*)}$ at complex values of $Q$. 
Similar remarks apply to $\tilde R_{\tau}$. One might note that in the NNLO 
case, where we can compare with the exact result, $\rho_{2}^{D(L)}$ is only a 
good approximation to the exact $\rho_{2}$ for $N_{f}\approx0$ or for large 
$N_{f}$. Hence one could 
conclude that the uncertainties in the basic approximation are such that the 
attempted improvement is not warranted. Nonetheless it would be 
very worrying if 
any of our conclusions for $R$, $R_{\tau}$ changed on including these extra 
terms. We hope to check this in a future work.

Another aspect of the resummations which requires elucidation is the way the 
resummed $\rho^{(L*)}$ effective charge beta-function is obtained by 
numerically 
inverting the P.V. regulated Borel integral representation of $D^{(L)}(a)$, as 
detailed in equations (35)-(37). From equation (30) we see that $\rho(D(Q))$ is 
directly related to the $Q$-evolution of the observable $D(Q)$, 
and is therefore 
of central physical importance in studying power corrections. One might then 
imagine defining
\be
\rho(D)=``Reg.''\int_{0}^{\infty}\mbox{d}z\,\mbox{e}^{-z/D}B[\rho](z)+
\rho_{NP}^{Reg}(D)
\;,
\ee
where $B[\rho]$ denotes the perturbatively defined Borel transform of $\rho$. 
This will contain singularities  at the same positions in the Borel plane as 
$D(a)$ \cite {ben2}, and to control the I${\rm R}_{l}$ infra-red renormalon 
singularities 
on the positive-$z$ axis the integral will have to be regulated, denoted 
``$Reg.$''. There will be an additional $\rho_{NP}^{Reg}(D)$ incorporating the 
power corrections ($e^{-1/D}$ terms) whose precise definition will 
depend on the 
chosen method of regulation \cite {grun3}.

If $B[\rho]$ is defined in the leading-$b$ approximation we can then ask if the 
first term in equation (42) with P.V. regulation exactly reproduces the 
$\rho^{(L*)}$ defined by numerically inverting the P.V. regulated $D^{(L)}(a)$.

This can be reduced to a simpler problem. Consider
\begin{eqnarray}
D(x)&\equiv&P.V.\int_{0}^{\infty}\mbox{d}z\,\mbox{e}^{-z/x}B[D](z)\nonumber\\
& & \\
a(x)&\equiv&P.V.\int_{0}^{\infty}\mbox{d}z\,\mbox{e}^{-z/x}B[a](z)\nonumber 
\end{eqnarray}
where $B[D]$ denotes the perturbatively defined Borel transform of $D(a)$, and 
$B[a]$ denotes the perturbatively defined Borel transform of the inverse 
function 
of $D(a)$, which can unambiguously be defined by formally transforming 
the coefficients of the power series $D(a)$. With these definitions 
one can then 
ask whether $D(a(x))=x$ exactly or whether there are additional $e^{-1/x}$ 
terms. Existing results on such problems are in short supply \cite {aub}, but 
unfinished work in progress \cite {barc} strongly suggests that the relation 
$D(a(x))=x$ 
does hold exactly. Unfortunately the result probably only holds for P.V. 
regulation. The pragmatic reason for studying this question  is that $B[\tilde 
D^{(L)}](z)$ is given by rather simple expressions as a sum of poles \cite 
{charles2}, 
whereas $B[\rho](z)$ will have an extremely complicated form. Hence it is 
impractical to construct $B[\rho](z)$ directly, and the numerical inversion 
route is the only possibility.

The properties of the function $\rho(x)$ fix the infra-red properties of the 
observable. For instance if $\rho(D^{*})=0$ then $D\rightarrow D^{*}$ as 
$Q\rightarrow 0$. It has been suggested \cite {stev2,stev3} that such infra-red 
freezing is 
supported by a wide body of indirect phenomenological evidence. In reference 
\cite {dok} the assumption of universal infra-red behaviour of an effective 
coupling 
$\alpha_{eff}(k)$ has been used to interrelate power corrections for different 
observables. It is interesting that for all the observables we have studied in 
this paper $D^{(L)}(a)$ has a maximum value, $D_{max}$, say. This means that 
$\rho^{(L)}(x)$ is undefined for $x>D_{max}$. If $\rho$ is to be defined in the 
infra-red this is presumably a signal that power corrections 
have to be included 
beyond a certain point. An interesting consistency check on this 
interpretation is that if only ultra-violet renormalon 
singularities are present then $D^{(L)}(a)$ does not have a maximum. In 
particular if, as is the case for $\tilde R$ and $\tilde R_{\tau}$, the UV 
singularities are single poles, then $D^{(L)}(a)$ increases monotonically and 
$\rho^{(L)}(x)$ will be defined for all $x$. The absence of IR renormalons is 
consistent with there being no power corrections, or at least they are not 
constrained by the large-order perturbative behaviour. We hope to take up this 
question of IR behaviour and constraining the form of power corrections in a 
later work.

A final underlying issue which needs further clarification is the explanation 
for the excellent performance of the leading-$b$ approximation itself. For all 
the cases where exact NNLO QCD calculations exist the leading-$b$ 
approximation not only gives exact results for perturbative coefficients and 
$\rho_{k}$ RS-invariants in the large-$N_{f}$ limit, but remarkably is also an 
excellent ($\sim$ 5\% level) approximation in the large-$N$ limit of a 
large number of colours. As pointed out unfortunately it may be a rather poor 
approximation in-between these extremes, for $N_{f}$=5, $N$=3, for instance. 
Although the sub-leading $N_{f}$-expansion coefficients are reproduced 
remarkably well ($\sim$ 20\% level).

A possible Feynman-diagrammatic explanation runs as follows. In the large-$N$ 
limit of QCD only planar diagrams contribute. `t Hooft has shown that if one 
restricts oneself to UV-finite planar diagrams perturbation theory converges 
\cite {hooft}. As far as perturbative estimates are concerned 
these diagrams can 
be 
discarded, therefore, since they do not contribute to $n!$ growth of the 
coefficients. The remaining UV-divergent planar diagrams will 
contain among them 
diagrams containing chains of gluon self-energy insertions and other structures 
which must be combined with renormalon diagrams with chains of internal fermion 
bubbles and other structures to produce a gauge-invariant contribution 
proportional to a power of $b$, using the pinch technique or background field 
method 
\cite 
{watson}. The planar diagrams of interest are 
those 
{\it not} involved in the construction of a gauge-invariant effective charge, 
therefore. The hope would be to understand why their contribution is `small'. 
This would not, unfortunately, explain why the RS-invariant effective charge 
beta-function coefficients are reproduced so well in the large-$N$ limit, since 
this involves a combination of perturbative coefficients and beta-function 
coefficients. The piece that is so well approximated in the large-$N$ limit is 
that which dominates in a Banks-Zaks expansion in 
$\varepsilon\equiv(\frac{11N}{2}-N_{f})$ around $\varepsilon=0$ \cite 
{bankszaks} where 
asymptotic freedom is lost. This has been studied in the context of IR freezing 
of observables since for suitably small $\varepsilon$ there is 
expected to be an 
IR fixed point \cite {stev3}.

\section{Conclusions}

In this paper we have proposed an improvement of the renormalon-inspired 
`leading-$b$' resummations of QCD perturbation theory which 
have been previously 
used by various authors [1--6] to assess the reliability of fixed-order 
perturbative predictions. Such resummations are RS-dependent under the full QCD 
RG transformations. To avoid this difficulty the strategy is to approximate the 
RS-invariant effective charge beta-function coefficients by retaining their 
`leading-$b$' part, which is completely determined by exact all-orders 
large-$N_{f}$ results. Fixed-order perturbative approximations in any RS can 
then be obtained from the approximated RS-invariants by using the exact QCD RG. 
If the exact NNLO invariant is known it can be included. In this way the 
resummation includes the exact NLO and NNLO perturbative coefficients in any RS.

The RS-invariant resummation was performed for the $e^{+}e^{-}$ $R$-ratio, 
$R_{\tau}$ the analogous decay ratio for the tau-lepton, and DIS sum rules. 
Comparison with fixed-order perturbation theory in the effective charge RS 
revealed impressive convergence to the resummed result for the $e^{+}e^{-}$ 
$R$-ratio at LEP/SLD energies, $Q=91$ GeV. As the value of $Q$ was reduced  
oscillatory behaviour of the fixed-order results above and below the resummed 
value was increasingly evident, reflecting the alternating-sign factorial 
growth of the perturbative coefficients resulting from the dominant 
U$\rm V_{1}$ 
renormalon singularity. Even at $Q=1.5$ GeV the resummed value was reasonably 
approximated until ninth order perturbation theory.

For $R_{\tau}(Q=m_{\tau})$, which is also U$\rm V_{1}$ dominated, there was 
also a satisfactory approximation to the resummed value, although with much 
larger oscillations than for $\tilde R$ at a comparable value of $Q$, and with 
an earlier breakdown of perturbation theory beyond fifth-order.
  
In contrast DIS sum rules which have an I$\rm R_{1}$ infra-red singularity 
exhibited much less satisfactory behaviour with successive orders moving 
steadily away from the resummed result, reflecting the fixed-sign factorial 
growth of the coefficients.

Using the difference between the exact NNLO EC approximation 
and the resummation 
to estimate the uncertainty with which $\Lambda_{\MS}$ could be determined 
indicates that for $\tilde R$ at $Q=91$GeV $\alpha_{s}(M_{Z})$ could be 
determined to three significant figures with ideal data.

For $\tilde R_{\tau}$ one concludes that $\delta\alpha_{s}(M_{Z})=0.001$ from 
the 
NNLO-resummed difference. This is a much smaller uncertainty than deduced by 
Neubert 
\cite {bert2} from a comparison with the naive RS-dependent leading-$b$ 
resummation. The 
RS-dependence means that the naive resummation is sensitive to the $\MS$ scheme 
$\alpha_{s}(m_{\tau})$ being assumed for the coupling. Other {\it a priori} 
reasonable 
choices would dramatically change the resummed result, and hence we would argue 
that this estimate of the uncertainty is too pessimistic.

We regard the impressive performance of fixed-order QCD perturbation theory for 
the UV-renormalon dominated quantities as the key result of this analysis.

Various technical issues  related to the resummation and possibilities for 
future developments were also discussed.

\section*{Acknowledgements}

We would like to thank David Barclay, Georges Grunberg, Jan Fischer, and Andrei 
Kataev for a number of stimulating discussions.

D.G.T gratefully acknowledges receipt of a P.P.A.R.C. U.K. Studentship.


\newpage
\begin{thebibliography}{99}

\bibitem{charles2} C.N.Lovett-Turner and C.J.Maxwell, Nucl.Phys. {\bf B452} 
(1995) 188.
\bibitem{benbraun1} M.Beneke and V.M.Braun, Phys.Lett. {\bf B348} (1995) 513.
\bibitem{ball} P.Ball, M.Beneke and V.M.Braun, Nucl.Phys. {\bf B452} (1995) 563.
\bibitem{bert1} M.Neubert, CERN-TH 7524/94 [hep-ph/9502264]; 
Phys.Rev. {\bf D51} 
(1995) 5924.
\bibitem{bert2} M.Neubert, CERN-TH/95-112 [hep-ph/9509432].
\bibitem{bert3} M.Girone and M.Neubert, CERN-TH/95-275 [hep-ph/9511392].
\bibitem{ben1} M.Beneke, Nucl.Phys. {\bf B405} (1993) 424.
\bibitem{broad} D.J.Broadhurst, Z.Phys. {\bf C58} (1993) 339.
\bibitem{broadkat} D.J.Broadhurst and A.L.Kataev, Phys.Lett. {\bf B315} (1993) 
179. 
\bibitem{benbraun2} M.Beneke and V.M.Braun, Nucl. Phys. {\bf B426} 
(1994) 301.
\bibitem{bert4} M.Neubert and C.Sachrajda, Nucl.Phys. {\bf B438} (1995) 235.
\bibitem{watson} N.J.Watson, Phys.Lett. {\bf B349} (1995) 155.
\bibitem{charles1} C.N.Lovett--Turner and C.J.Maxwell, Nucl. Phys. 
{\bf B432} (1994) 147.
\bibitem{kat} K.G.Chetyrkin, A.L.Kataev and F.V.Tkachov, Phys.Lett. {\bf B85} 
(1979) 277; M.Dine and J.Sapirstein, Phys.Rev.Lett. {\bf 43} (1979) 668; 
W.Celmaster and R.J.Gonsalves, Phys.Lett. {\bf B44} (1980) 560.
\bibitem{gorish1} S.G.Gorishny,~A.L.Kataev~and~S.A.Larin, Phys.Lett. {\bf B259} 
(1991) 
144;\\
L.R.Surguladze and M.A.Samuel, Phys.Rev.Lett. {\bf 66} (1991) 560; {\bf 66} 
(1991) 
2416 
(E).
\bibitem{chyla} J.Chyla, Phys.Lett. {\bf B356} (1995) 341.
\bibitem{gorish2} S.G.Gorishny and S.A.Larin, Phys.Lett. {\bf B172} (1986) 109; 
E.B.Zijlstra and W. van Neerven, Phys.Lett. {\bf B297} (1992) 377.
\bibitem{larin} S.A.Larin and J.A.M.Vermaseren, Phys.Lett. {\bf B259} (1991) 
345.
\bibitem{reader} D.T.Barclay, C.J.Maxwell and M.T.Reader, Phys.Rev {\bf D49} 
(1994) 3480.
\bibitem{stev1} P.M.Stevenson, Phys.Rev. {\bf D23} (1981) 2916.
\bibitem{grun1} G.Grunberg, Phys.Lett. {\bf B95} (1980) 70; Phys.Rev. {\bf D29} 
(1984) 2315.
\bibitem{grun2} G.Grunberg, Phys.Rev. {\bf D46} (1992) 2228.
\bibitem{ben2} M.Beneke, Phys.Lett. {\bf B307} (1993) 154.
\bibitem{grun3} G.Grunberg, Phys.Lett. {\bf B325} (1994) 441.
\bibitem{lep} LEP Collaborations Joint Report. CERN-PPE/93-157 (1993).
\bibitem{marshall} R.Marshall, Z.Phys. {\bf C43} (1989) 595.
\bibitem{rahal} G.Rahal-Callot, to appear in Proc.Int.Europhysics.Conf. on High 
Energy Physics, Brussels, Belgium, September 1995.
\bibitem{wasser} S.R.Wasserbaech, to appear in Proc.Workshop on Tau/Charm 
Factory, Argonne, Illinois, June 1995.
\bibitem{ccfr} CCFR-NuTeV collaboration. Contribution to 30th Rencontres de 
Moriond: QCD and High Energy Hadronic Interactions, Meribel Les Allues, France, 
19-25 March 1995. [hep-ex/9506010].
\bibitem{braun} V.M.Braun and A.V.Kolesnichenko, Nucl.Phys. {\bf B283} (1987) 
723.
\bibitem{pich} F.Le Diberder and A.Pich, Phys.Lett. {\bf B286} (1992) 147; F.Le 
Diberder and A.Pich, Phys.Lett. {\bf B289} (1992) 165.
\bibitem{racz1} P.Raczka and A.Szymacha, Institute of Physics, Warsaw 
University, preprint IFT/1/95, [hep-ph/9602245].
\bibitem{racz2} P.Raczka and A.Szymacha, Institute of Physics, Warsaw 
University, preprint IFT/13/94, [hep-ph/9412236].
\bibitem{aub} G.Auberson and G.Mennessier, J.Math.Phys. {\bf 22} (1981) 2472;
Commun.Math.Phys. {\bf 100} (1985) 439.
\bibitem{barc} D.T.Barclay, private communication.
\bibitem{stev2} A.C.Mattingly and P.M.Stevenson, Phys.Rev. {\bf D49} (1994) 437.
\bibitem{stev3} P.M.Stevenson, Phys.Lett {\bf B331} (1994) 187.
\bibitem{dok} Yu.L.Dokshitzer, G.Marchesini and B.R.Webber. CERN-preprint 
CERN-TH/95-281 [hep-ph/9512336].
\bibitem{hooft} G.`t Hooft, Commun.Math.Phys. {\bf 86} (1982) 449.
\bibitem{bankszaks} T.Banks and A.Zaks, Nucl. Phys. {\bf B206} (1982) 23.
\end{thebibliography}
\end{document}